\documentclass[graybox]{svmult}
\usepackage{amssymb,mathrsfs,amsmath,amsfonts,graphicx,pdfpages,todonotes,bbold,float}
\usepackage{caption}
\usepackage{subcaption}

\usepackage{xcolor}
\usepackage{framed,color}

\definecolor{shadecolor}{rgb}{1,0.8,0.3}
\definecolor{myurlcolor}{rgb}{0.5,0,0}
\definecolor{mycitecolor}{rgb}{0,0,0.7}
\definecolor{myrefcolor}{rgb}{0,0,0.7}
\definecolor{hyperrefcolor}{rgb}{0.5,0,0}

\usepackage{hyperref}
\hypersetup{
	colorlinks,
	linkcolor={hyperrefcolor},
	citecolor={mycitecolor},
	urlcolor={hyperrefcolor}
}

\newcommand{\namedset}[1]{\mathbb{#1}}
\newcommand{\R}{\namedset{R}}

\newcommand{\namedcat}[1]{\mathsf{#1}}
\newcommand{\Set}{\namedcat{Set}}

\newcommand{\G}{\namedcat{G}}
\renewcommand{\H}{\namedcat{H}}

\newcommand{\Dynam}{\namedcat{Dynam}}
\newcommand{\StockFlow}{\namedcat{StockFlow}}
\newcommand{\CausalLoop}{\namedcat{CausalLoop}}
\newcommand{\SystemStructure}{\namedcat{SystemStructure}}

\newcommand{\X}{\namedcat{X}}

\renewcommand{\S}{\mathrm{S}}
\newcommand{\F}{\mathrm{F}}
\newcommand{\V}{\mathrm{V}}
\newcommand{\SV}{\mathrm{SV}}
\renewcommand{\I}{\mathrm{I}}
\renewcommand{\O}{\mathrm{O}}
\newcommand{\LS}{\mathrm{LS}}
\newcommand{\LV}{\mathrm{LV}}
\newcommand{\LSV}{\mathrm{LSV}}

\newcommand{\is}{\mathrm{is}}
\newcommand{\ifn}{\mathrm{ifn}}
\newcommand{\os}{\mathrm{os}}
\newcommand{\ofn}{\mathrm{ofn}}
\newcommand{\lss}{\mathrm{lss}}
\newcommand{\lssv}{\mathrm{lssv}}
\newcommand{\lsvv}{\mathrm{lsvv}}
\newcommand{\fv}{\mathrm{fv}}
\newcommand{\lvs}{\mathrm{lvs}}
\newcommand{\lvv}{\mathrm{lvv}}
\newcommand{\lsvsv}{\mathrm{lsvsv}}

\newcommand{\CL}{\mathrm{CL}}
\renewcommand{\E}{\mathrm{E}}
\newcommand{\Node}{\mathrm{N}}

\newcommand{\N}{\mathrm{N}}
\newcommand{\NS}{\mathrm{NS}}
\newcommand{\NI}{\mathrm{NI}}

\makeatletter
\newcommand*{\relrelbarsep}{.386ex}
\newcommand*{\relrelbar}{%
  \mathrel{%
    \mathpalette\@relrelbar\relrelbarsep
  }%
}
\newcommand*{\@relrelbar}[2]{%
  \raise#2\hbox to 0pt{$\m@th#1\relbar$\hss}%
  \lower#2\hbox{$\m@th#1\relbar$}%
}
\providecommand*{\rightrightarrowsfill@}{%
  \arrowfill@\relrelbar\relrelbar\rightrightarrows
}
\providecommand*{\leftleftarrowsfill@}{%
  \arrowfill@\leftleftarrows\relrelbar\relrelbar
}
\providecommand*{\xrightrightarrows}[2][]{%
  \ext@arrow 0359\rightrightarrowsfill@{#1}{#2}%
}
\providecommand*{\xleftleftarrows}[2][]{%
  \ext@arrow 3095\leftleftarrowsfill@{#1}{#2}%
}
\makeatother

\usepackage{enumerate}

\newcommand{\maps}{\colon}

\usepackage{type1cm}        % activate if the above 3 fonts are
\usepackage{makeidx}         % allows index generation
\usepackage{graphicx}        % standard LaTeX graphics tool
\usepackage{multicol}        % used for the two-column index
\usepackage[bottom]{footmisc}% places footnotes at page bottom

\usepackage{newtxtext}       % 
\usepackage[varvw]{newtxmath}       % selects Times Roman as basic font

\makeindex             % used for the subject index
\newcommand{\tx}[1]{\textrm{#1}}

\begin{document}

\title*{A Categorical Framework for Modeling with Stock and Flow Diagrams}
\author{John C.\ Baez, Xiaoyan Li, Sophie Libkind, Nathaniel D.\ Osgood, Eric Redekopp}
\institute{John C.\ Baez \at Department of Mathematics, University of California, Riverside CA 92507, USA, \email{baez@math.ucr.edu} 
\and Xiaoyan Li \at Department of Computer Science, University of Saskatchewan, 
%176 Thorvaldson Bldg, 110 Science Place, Saskatoon,
Saskatoon, SK, S7N 5C9, Canada, \email{xiaoyan.li@usask.ca}
\and Sophie Libkind \at Department of Mathematics, Stanford University, Palo Alto, CA, 94305, USA, \email{slibkind@stanford.edu}
%\and Long Pham \at Department of Computer Science, University of Saskatchewan, 
%% 176 Thorvaldson Bldg, 110 Science Place,
%Saskatoon, SK, S7N 5C9, Canada, \email{ngp143@mail.usask.ca}
\and Nathaniel D.\ Osgood \at Department of Computer Science, University of Saskatchewan, 
% 176 Thorvaldson Bldg, 110 Science Place,
Saskatoon, SK, S7N 5C9, Canada, \email{nathaniel.osgood@usask.ca}
\and Eric Redekopp \at Department of Computer Science, University of Saskatchewan, 
%176 Thorvaldson Bldg, 110 Science Place, Saskatoon, 
Saskatoon, SK, S7N 5C9, Canada, \email{eric.redekopp@usask.ca}}
 

\maketitle

\abstract{
Stock and flow diagrams are already an important tool in epidemiology, but category theory lets us go further and treat these diagrams as mathematical entities in their own right.  In this chapter we use communicable disease models created with our software, StockFlow.jl, to explain the benefits of the categorical approach. We first explain the category of stock-flow diagrams and note the clear separation between the syntax of these diagrams and their semantics, demonstrating three examples of semantics already implemented in the software: ODEs, causal loop diagrams, and system structure diagrams. We then turn to two methods for building large stock-flow diagrams from smaller ones in a modular fashion: composition and stratification. Finally, we introduce the open-source ModelCollab software for diagram-based collaborative modeling. The graphical user interface of this web-based software lets modelers take advantage of the ideas discussed here without any knowledge of their categorical foundations.}

\section{Introduction}
\label{section: introduction}
%1.1 background

Mathematical modeling of infectious disease at scale is important, but challenging.  There are many benefits to the modeling process extending from taking diagrams as mathematical formalisms in their own right with the help of category theory.  Stock and flow diagrams are widely used in infectious disease modeling, so we illustrate this point using these.  However, rather than focusing on the underlying mathematics, we informally use communicable disease examples created with our software, called StockFlow.jl \cite{AlgebraicStockFlow}, to explain the benefits of the categorical framework. Readers interested in the mathematical details may refer to our earlier paper \cite{baez2022compositional}.

%Stock and Flow diagrams are a formalism used to depict model structure in many infectious disease modeling projects. 

%1.2 also introducing system structure diagrams, stock and flow diagrams, causal loop diagrams

Many compartmental modelers regard diagrams offering a visual characterization of structure---e.g., susceptible, infective and recovered stocks and transitions between them---as broadly accessible but informal steps towards a mathematically rigorous formulation in terms of ordinary differential equations (ODEs). However,  ODEs are typically opaque to non-modelers---including the interdisciplinary members of the teams that typically are required for impactful models. By contrast, the System Dynamics modeling tradition places a premium on engagement with stakeholders \cite{hovmandCBSD}, and offers a modeling approach centered around diagrams. This approach commonly proceeds in a manner that depicts model structure using successively more detailed models. The process starts with a ``causal loop diagram'' illustrating causal connections and feedback loops (Figure \ref{fig:seir_causal_loop}).  It then often proceeds to a ``system structure diagram'', which distinguishes stocks from flows but still lacks quantitative information.  The next step is to construct a stock and flow diagram---or henceforth, ``stock-flow diagram'' (Figure \ref{fig:SEIR_stockflow}).  This diagram is visually identical to the system structure diagram, but it also includes formulae, values for parameters, and initial values of stocks.

The stock--flow diagram is treated as the durable end result of this modeling process, since it uniquely specifies a system of first-order ODEs.  System Dynamics modeling typically then alternates between assessing scenario outcomes resulting from numerically integrating the ODEs, performing other analyses (e.g., identifying location or stability of equilibria), and elaborating the stock-flow diagram---such as by adding elements to it, often borrowed from other models, or ``stratifying'' it by breaking large stocks (compartments) into smaller ones that differ in some characteristics.

While each of the types of diagrams in the System Dynamics tradition is recommended by visual accessibility, development of models using the traditional approach suffers from a number of practical shortcomings.

\begin{enumerate}
  \item{\textbf{Monolithic models}: Stock-flow models are traditionally built up in a monolithic fashion, leading ultimately to a single large piece of code.  In larger models, this inhibits independent simultaneous work by multiple modelers.  Lack of model modularity further prevents effective reuse of particular model elements. If elements of other models are used, they are commonly copy-and-pasted into the developing model, with the source and destination then evolving independently. Such separation can lead to a proliferation of conceptually overlapping models in which a single conceptual change (e.g., addition of a new asymptomatic infective compartment) requires corresponding updates in several successive models.}

  \item{\textbf{Curse of stratification dimensionality}: While stratification is a key tool for representing heterogeneity and multiple lines of progression in compartmental models, stratification commonly requires modifications across the breadth of a model---stocks, flows, derived quantities, and many parameters. When that stratification involves multiple dimensions of heterogeneity, it can lead to a proliferation of terms in the ODEs. For example, rendering a model characterizing both COVID-19 into a model also characterizes influenza would require that each COVID-19 state to be replicated for each stage in the natural history of influenza. Represented visually, this stratification leads to a multi-dimensional lattice, commonly with progression proceeding along several dimensions of the lattice. Because of the unwieldy character of the diagram, much of the structure of the model is obscured. Adding, removing, or otherwise changing dimensions of heterogeneity routinely leads to pervasive changes across the model.}
  \item{\textbf{Privileging ODE semantics}: The structure of causal loop diagrams, system structure diagrams and stock-flow diagrams characterizes general state and accumulations, transitions and posited causal relations---including induced feedbacks---amongst variables. Nothing about such a characterization restricts its meaning to ordinary differential equations; indeed, many other interpretations and uses of these diagrams are possible. However, existing software privileges an ODE interpretation for stock-flow diagrams, while sometimes allowing for secondary analyses in \textit{ad hoc} way---for example, identifying causal loops associated with the model, or verifying dimensional homogeneity in dimensionally annotated models.  Conducting other sorts of analyses---such as computation of eigenvalue elasticities or loop gains, analysis as a stochastic transition system, or other methods such as particle filtering \cite{Arulampalam2002-xj,li2018applyingMeasles,OsgoodLiuPF2014,  safarishahrbijari2017predictive}, particle MCMC \cite{andrieuDoucet2010PMCMC,opioids_li2018illuminating,SSC_OsgoodEng2022} or Kalman filtered \cite{gelb1974applied, WinchellKalman} systems---typically requires bespoke software for reading, representing and analyzing stock-flow models.}
  \item{\textbf{Divergence of model representations}: Although the evolution from causal loop diagrams to system structure diagrams to stock-flow models is one of successive elaboration and informational enrichment, existing representations treat these as entirely separate characterizations and fail to capture the logical relationships between them. Such fragmentation commonly induces inconsistent evolution. Indeed, in many projects, the evolution of stock-flow diagrams renders the earlier, more abstract formulations obsolete, and the focus henceforth rests on the stock-flow diagrams.}
\end{enumerate}

What is less widely appreciated is that beyond their visual transparency and capacity to be lent a clear ODE semantics, both stock-flow diagrams themselves and their more abstract cousins possess a precise mathematical structure---a corresponding grammar, as it were.  This algebraic structure, called the ``syntax'' of stock-flow diagrams, can be characterized using the tools of a branch of mathematics called category theory \cite{SevenSketches, leinster}. Formalizing the syntax of stock-flow diagrams lends precise meaning to the process of ``composing'' such models (building them out of smaller parts), stratifying them, and other operations. Explicitly characterizing the syntax in software also allows for diagrams to be represented, manipulated, composed, transformed, and flexibly analyzed in software that implements the underlying mathematics.

Formalizing the mathematics of diagram-based models using category theory and capturing it in software offers manifold benefits.  This paper discusses and demonstrates just a few:
\begin{enumerate}
  \item{\textbf{Separation of syntax and semantics.}  Category theory gives tools to separate the formal structure, or ``syntax'', of diagram-based models from the uses to which they are put, or ``semantics''.  This separation permits great flexibility in applying different semantics to the same model.  With appropriate software design, this decoupling can allow the same software to support a diverse array of analyses, which can be supplemented over time.}
  \item{\textbf{Reuse of structure.}  The approaches explored here provide a structured way to build complex diagrams by composing small reusable pieces.  With software support, modeling frameworks can allow for saving models and retrieving them for reuse as parts of many different models. For example, a diagram representing contact tracing processes can be reused across diagrams addressing different pathogens.}
  \item{\textbf{Modular stratification.} A categorical foundation further supports a structured way to build stratified dynamical systems out of modular, reusable, largely orthogonal pieces. In contrast to the global changes commonly required to a diagram and the curse of dimensionality that traditionally arises when stratifying a diagram, categorically-founded stratification methods allow for crisply characterizing a stratified diagram as built from simpler diagrams, one for each heterogeneity or progression dimension.}
\end{enumerate}

The balance of the chapter is structured as follows. Section \ref{section: closed stock flow} describes the categorical formulation of stock-flow diagrams. Section \ref{section: semantics} explains how the categorical approach allows for decoupling the syntax and semantics of such diagrams. Section \ref{section: open stock flow} introduces the hallmark of the categorical approach: the ability to build larger stock-flow diagrams from smaller pieces by composition.  Section \ref{section: stratification} discusses another key application of the categorical approach: stratification.  Section \ref{section: ModelCollab} introduces ModelCollab: a web-based graphical user interface that allows users to build and run stock-flow models using the category-theoretic ideas we have introduced without requiring expertise in the mathematics.  Finally, in Section \ref{section: conclusion} we close with some reflections on the significance and evolution of the methods described here. The code for examples in this paper can be found in the Appendix\footnote{The code can also be found in the GitHub repository \url{https://github.com/Xiaoyan-Li/applicationStockFlowMFPH}.}.

\section{The Syntax of Stock-Flow Diagrams}
\label{section: closed stock flow}

In this section, we illustrate the categorical method of representing stock-flow diagrams, which we can think of as characterizing the syntax of such diagrams. Section 3 of our previous work \cite{baez2022compositional} introduced the mathematics underlying stock-flow diagrams. In this paper, we show how to encode an instance of a familiar stock-flow diagram---namely, the Susceptible-Exposure-Infectious-Recovered or ``SEIR'' model with an open population \cite{anderson1992infectious}---using the categorical method. 

\begin{figure}[H]
    \centering
    \includegraphics[width=0.9\columnwidth]{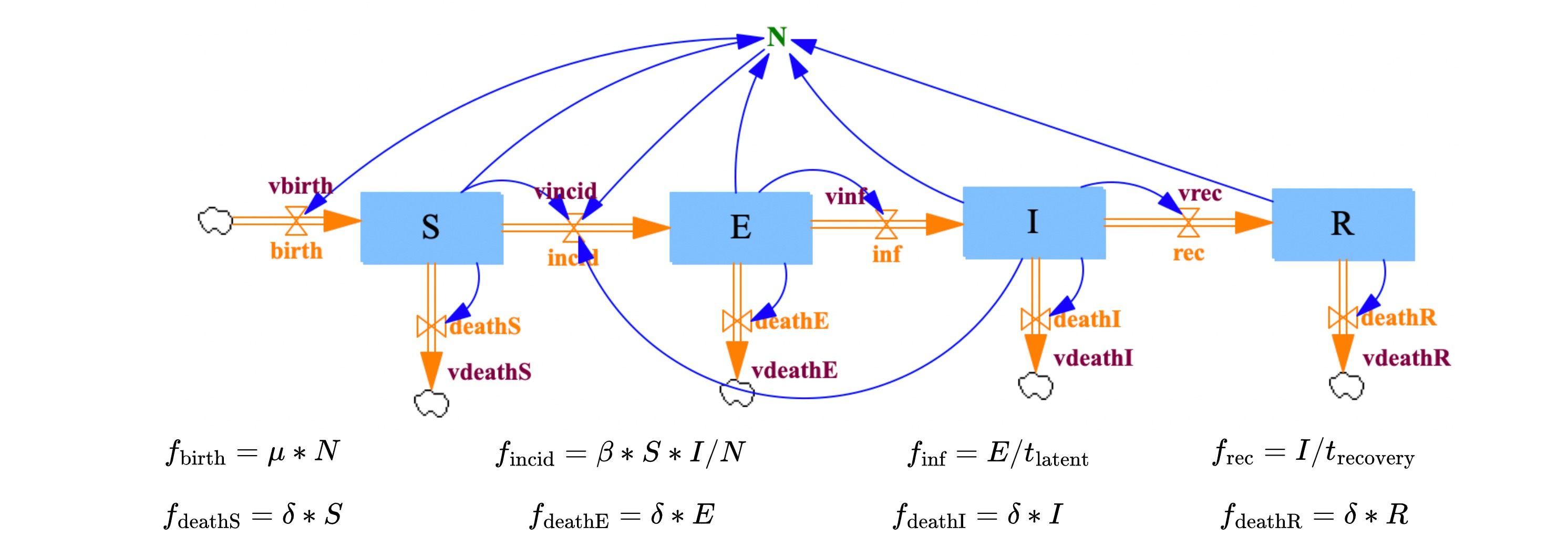}
    \caption{The stock-flow diagram of the open population SEIR model}
    \label{fig:SEIR_stockflow}
\end{figure}

Figure \ref{fig:SEIR_stockflow} shows the stock-flow diagram for the SEIR model.  The blue boxes labeled S, E, I and R are ``stocks''.  The letter N is a ``sum variable'', and there are blue ``links'' to it from the stocks on which it depends.   The orange arrows are ``flows''.  Note that some flows go between stocks, while others enter a stock from outside the model, or leave a stock to go outside the model. The latter two cases are indicated with small ``clouds''.   

There are one or more blue links to each flow from the stocks and/or sum variables on which it depends.  For each flow there is also a ``flow variable'' drawn in purple, which indicates the rate of that flow.  Each flow variable is determined by some function of the stocks and/or sum variables that are connected to that flow variable by blue links.   These functions are defined below the diagram in Figure \ref{fig:SEIR_stockflow}.   For example, the flow describing infection, called $\textrm{inf}$, has a flow variable $\textrm{vinf}$ determined by the function $f_{\textrm{inf}}$.

\begin{figure}[H]
   \sidecaption
    \includegraphics[width=0.5\columnwidth]{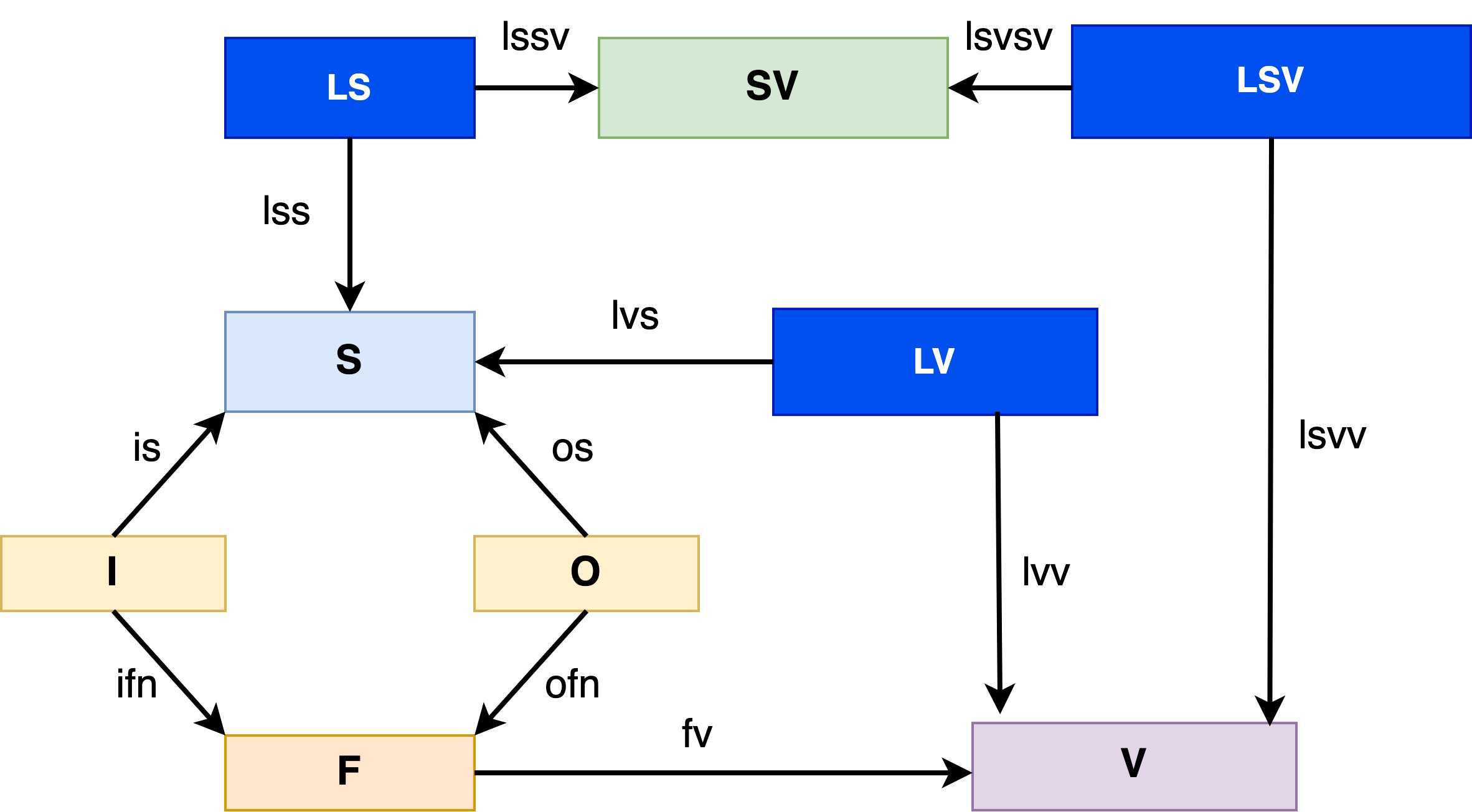}
    \caption{The schema for stock-flow diagrams}
    \label{fig:schema_stockflow}
\end{figure}

The SEIR model is just one example of a stock-flow diagram.  To formalize their general structure we use the so-called ``schema" for stock-flow diagrams, shown in Figure \ref{fig:schema_stockflow}.  It consists of boxes, called ``objects'', and arrows between boxes, called ``morphisms''.  In an ``instance'' of this schema, we choose a set for each object and a function between such sets for each morphism.  Other types of diagrams, such as causal loop diagrams, have their own schemas (see Section \ref{subsection: causal loop diagrams}), and in fact there is a general theory of schemas \cite{SevenSketches, spivak2014category}.   For now we consider only the schema for stock-flow diagrams.  In this schema:

\begin{enumerate}
  \item{The objects $\S$ and $\F$ represent the stocks and flows, respectively.}
  \item{The objects $\I$ and $\O$ represent the inflows and outflows.  The morphisms $\S \xleftarrow{\is} \I \xrightarrow{\ifn} \F$ are used to describe which stocks are downstream of a given flow, while the morphisms $\S \xleftarrow{\os} \O \xrightarrow{\ofn} \F$ are used to describe which stocks are upstream of a given flow. This structure allow for flows that go between two stocks, but also flows that enter a stock from outside the model or leave a stock and go outside the model.}
  \item{The object $\V$ represents ``auxiliary variables'', sometimes termed ``dynamic variables''.  These are variables whose value is an instantaneous function of the current model state. The morphism $\F \xrightarrow{\fv} \V$ indicates that the rate of each flow depends on one auxiliary variable. This relation is usually left implicit, not drawn, in the stock-flow diagram.}
  \item{The object $\LV$ represents ``variable links'': that is, links from stocks to auxiliary variables.  The morphisms $\S \xleftarrow{\lvs} \LV \xrightarrow{\lvv} \V$ indicate that any variable link goes from a stock to an auxiliary variable.}
  \item{The objects $\SV$ and $\LS$ represent ``sum variables'' and ``sum links''.   Sum variables are a special type of auxiliary variable introduced in \cite{baez2022compositional} to make composing stock-flow diagrams easier.  A sum variable is simply the sum of the stocks linked to them by sum links.  The arrows $\S \xleftarrow{\lss} \LS \xrightarrow{\lssv} \SV$ indicate that any sum link goes from a stock to a sum variable.}
   \item{The object $\LSV$ represents ``sum variable links'': that is, links from sum variables to auxiliary variables.  The arrows $\SV \xleftarrow{\lsvsv} \LSV \xrightarrow{\lsvv} \V$
   indicate that any sum variable link goes from a sum variable to an auxiliary variable.}
\end{enumerate}

An ``instance'' \(G\) of a schema assigns a finite set \(G(\mathsf{X})\) to each object \(\mathsf{X}\) of the schema and a function \(G(a) \maps G(\mathsf{X}) \to G(\mathsf{Y})\) to each morphism \(\mathsf{X} \xrightarrow{a} \mathsf{Y}\) of the schema.   So, an instance of the schema for stock-flow diagrams consists of:

\begin{enumerate}
  \item{A finite set of stocks $G(\S)$ and a finite set of flows $G(\F)$.}
  \item{A finite set of inflows $G(\I)$, a finite set of outflows $G(\O)$, and functions 
  \[ G(\is) \maps G(\I) \to G(\S),  \qquad G(\ifn) \maps G(\I) \to G(\F) \]
  \[ G(\os) \maps G(\O) \to G(\S), \qquad G(\ofn) \maps G(\O) \to G(\F).\]
  }
\item{A finite set of auxiliary variables $G(\V)$ and a function 
\[ G(\fv) \maps G(\F) \to G(\V).\]
}
\item{A finite set of variable links $G(\LV)$ and functions
\[  G(\lvs) \maps G(\LV) \to G(\S), \qquad 
G(\lvv) \maps G(\LV) \to G(\V).\]
}
\item{A finite set of sum variables $G(\SV)$, a finite set of sum links $G(\LS)$, and functions
\[  G(\lss) \maps G(\LS) \to G(\S), \qquad
G(\lssv) \maps G(\LS) \to G(\SV) .\]
}
\item{A finite set of sum variable links $G(\LSV)$ and functions
\[   G(\lsvsv) \maps G(\LSV) \to G(\SV), 
\qquad G(\lsvv) \maps G(\LSV) \to G(\V).\]}
\end{enumerate}
A ``stock-flow diagram'' is a pair $(G,\phi)$ consisting of an instance $G$ and, for each auxiliary variable $v$, a continuous function $\phi_v \maps \R^{G(\lvv)^{-1}(v)} \times \R^{G(\lsvv)^{-1}(v)} \to \R$.  In Section~\ref{subsection: ODEs} we explain how in the ODE semantics for stock-flow diagrams, the function $\phi_v$ specifies how the value of the variable $v$ depends on the stocks and sum variables that link to it.   

Given an inflow $i \in G(\I)$ we say the stock $G(\is)(i)$ is ``downstream'' from the flow $G(\ifn)(i)$.  Similarly, given an outflow $o \in G(\O)$ we say the stock $G(\os)(o)$ is ``upstream'' from the flow $G(\ofn)(o)$.   In practice, we only want stock-flow diagrams where the functions $G(\ifn)$ and $G(\ofn)$ are injective. This constraint ensures that each flow has at most one downstream stock and at most one upstream stock.  Also in practice we attach to each stock, flow, auxiliary variable or sum variable an ``attribute'' which serves as its name.  This naming relies on the theory of attributes developed by Patterson, Lynch and Fairbanks \cite{patterson-lynch-fairbanks2021}.

\begin{figure}[t]
    \centering
    \includegraphics[width=1.0\columnwidth]{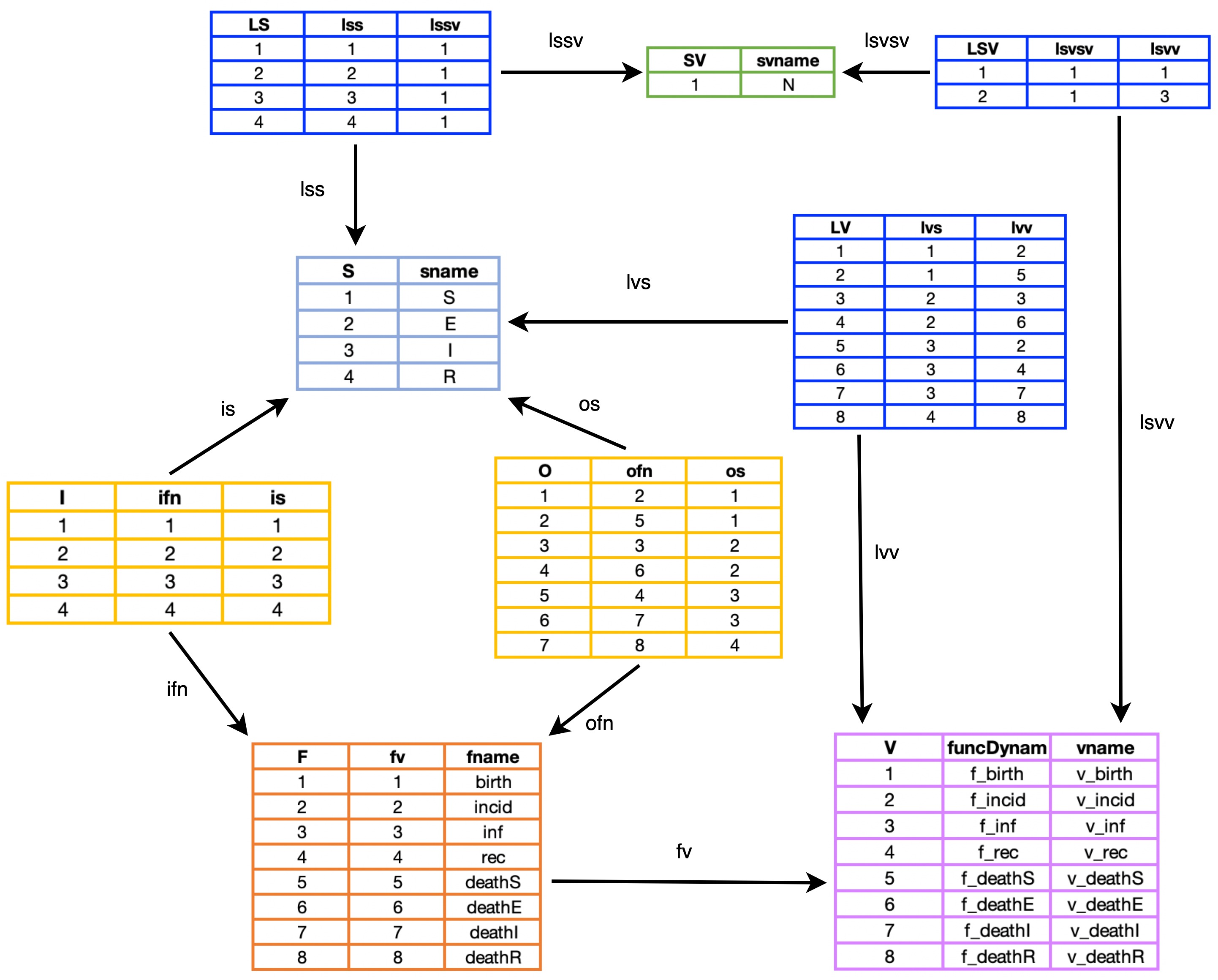}
    \caption{The categorical database structure representing the stock-flow diagram associated with the SEIR model}
    \label{fig:schema_database_stockflow}
\end{figure}

When a stock-flow diagram is implemented in software, it is encoded as a categorical database \cite{patterson-lynch-fairbanks2021}.  Figure \ref{fig:schema_database_stockflow} shows the categorical database representing the stock-flow diagram for the SEIR model shown in Figure \ref{fig:SEIR_stockflow}.  In this categorical database, each object $\X$ in the schema for stock-flow diagrams is represented by a database table. The row indices within this table consist of the elements of the set $G(\X)$ associated to the object $\X$.  Thus, if the set $G(\X)$ has $n$ elements, then the table has $n$ rows. In database parlance, each such row is associated with a ``primary key'' given in the first column of that table.  For example, since there are four stocks in the SEIR model, the object $\S$ in the schema for stock-flow diagrams maps to the set $\{1,2,3,4\}$, and the table for the object $\S$ has four rows numbered $1, 2, 3, 4$.   By contrast, the table for the object $\F$ has eight rows, reflecting the fact that there are eight flows in the SEIR model. 

The table for any object $\X$ has one column for each morphism coming out of $\X$, which describes the function associated to that morphism.   For example, the table for the object $\LV$ has one column ``$\lvs$'' describing the function $G(\lvs) \maps G(\LV) \to G(\S)$ mapping each variable link to a stock (as a ``foreign key'' giving the key of that stock in the ``S'' table), and one column ``$\lvv$'' describing the function $G(\lvv) \maps G(\LV) \to G(\V)$ mapping each variable link to a variable (similarly specified by a foreign key).  Besides this, the tables for $\S, \F, \V$ and $\SV$ have an extra column giving names for the stocks, flows and variables and sum variables, rather than foreign keys.  Technically, these names are handled using the theory of attributes mentioned above.

This capacity to encode stock-flow diagrams in a mathematically precise and transparent fashion confers diverse benefits. To list a few, these include the capacity to compose such diagrams (see Section \ref{section: open stock flow}), to soundly transform such diagrams for optimization, and to parallelize them. But one of the most foundational benefits is the capacity to perform different types of analysis on such diagrams---that is, to interpret a given diagram using various choices of ``semantics''. The next section discusses that benefit in greater detail.

\section{The Semantics of Stock-Flow Diagrams}
\label{section: semantics}

The capacity to interpret stock-flow diagrams in different ways is achieved by separating the syntax from the semantics of such diagrams.  Matters of syntax concern the forms that such diagrams can take.   In particular, the rules governing what counts as a legitimate stock-flow diagram---what can connect to what---are largely specified by the schema discussed in the previous section.  But the syntax of stock-flow diagrams is distinct from the interpretation given to these diagrams.  A given stock-flow diagram can be interpreted in different ways, each lending that diagram some meaning. Some of these interpretations involve dynamic simulations that solve equations specified by the diagram, whilst others describe static features of the diagram---for example, extracting algebraic equations specifying the equilibria of the system in terms of the parameters.  

In this section, we introduce three choices of semantics for stock-flow diagrams that have been implemented in StockFlow.jl \cite{baez2022compositional, AlgebraicStockFlow}: ordinary differential equations, and a pair of semantics that extract from a stock-flow diagram the associated causal loop diagram and system structure diagram.

\subsection{ODEs (Ordinary Differential Equations)}\label{subsection: ODEs}

A stock-flow diagram is traditionally used to represent a continuous-time, continuous-state dynamical system \cite{sterman2000business}: a system of ODEs that describes the evolution of each real-valued stock. While this addresses an important subclass of dynamical systems, this convention has needlessly restricted analysis potential and crimped the flexibility of supporting software by privileging a single interpretation of the syntax of the stock-flow diagrams. In this project, we mathematically decouple the choice of the stock-flow diagram (the syntax) from the choice of its interpretation (the semantics). As we shall see in this subsection, this approach readily supports the traditional interpretation of a stock-flow diagram in terms of ODEs. But as subsequent subsections illustrate, this interpretation is no longer required, or even privileged.

The decoupling of syntax and semantics afforded by the categorical approach is achieved by use of a structure-preserving map, called a ``functor'', sending each stock-flow diagram to its interpretation, or meaning.   The choice of different such functors allows for different interpretations. 

``Functorial semantics''---the idea of treating semantics as a functor---goes back to Lawvere's work \cite{lawvere1963functorial} in the early 1960s.  It has grown into a powerful method for specifying and analyzing the semantics of programming languages.   By now, it has also been applied to many diagrammatic modeling languages, including Petri nets, electrical circuit diagrams and chemical reaction networks, and others \cite{baezcourser2020,baez-courser-vasilakopoulou2022,fongthesis}.  Thus, the time is ripe for applying functorial semantics to stock--flow diagrams.

To do this, we need to define a \emph{category} of stock-flow diagrams, which in essence means precisely defining not only these diagrams, as we have done above, but also maps between them.  Using the methods of our previous paper \cite{baez2022compositional} we can define such a category, called $\StockFlow$, and also a functor
\[
\begin{tikzpicture}[baseline=(current  bounding  box.center),scale=1.5]
\node (A) at (0,0) {$\StockFlow$};
\node (B) at (2,0) 
{$\Dynam$};
\path[->,font=\scriptsize]
(A) edge node [above] {$v$} (B);
\end{tikzpicture}
\]
from this category to a category of ODEs and maps between those.  Here, for simplicity, we simply explain how this functor sends any stock-flow diagram to an ODE.

For any stock $s\in G(\S)$, the set of its inflows is $G(\is)^{-1}(s)$. Thus, this stock is downstream from precisely the flows in $G(\ifn)(G(\is)^{-1}(s))$. Similarly, this stock is upstream from precisely the flows in $G(\ofn)(G(\os)^{-1}(s))$. We denote as $\phi_v$ the continuous function describing the value of each auxiliary variable $v\in G(\V)$ as a function of the stocks and sum variables linked to it, so $\phi_v \maps \R^{G(\lvv)^{-1}(v)} \times \R^{G(\lsvv)^{-1}(v)} \to \R$. Then, the function describing the rate of the flow $f \in G(\F)$ is $\phi_{(G(\fv)(f))}$. The ordinary differential equation of the stock $s$ can then be defined as follows:

\begin{equation} \label{ODE_s} 
\dot{s}=\sum_{f \in G(\ifn)(G(\is)^{-1}(s))}\phi_{(G(\fv)(f))}-\sum_{f \in G(\ofn)(G(\os)^{-1}(s))}\phi_{(G(\fv)(f))},
\end{equation}
 Furthermore the value of each sum auxiliary variable is the sum of the values of the stocks it links to. For example, the value of the sum variable $N$ is defined as $N=S+E+I+R$ in the SEIR model.  

We can easily understand the functor mapping from a stock-flow diagram to the ODEs by the slogan: \textit{the time derivative of the value of each stock equals the sum of its inflows minus the sum of its outflows}.

For example, the SEIR stock-flow diagram (Figure \ref{fig:SEIR_stockflow}) gives the following set of ODEs:
\begin{equation} \label{ODEs_SEIR} 
 \begin{split}
 \dot{S}&=\mu N- \beta \frac{I}{N}S-\delta S\\
 \dot{E}&=\beta \frac{I}{N}S-\frac{E}{t_{\text{latent}}} -\delta E\\
 \dot{I}&=\frac{E}{t_{\text{latent}}}-\frac{I}{t_{\text{recovery}}}-\delta I\\
 \dot{R}&=\frac{I}{t_{\text{recovery}}}-\delta R
 \end{split}
\end{equation}
The solutions can be calculated and plotted out directly using the StockFlow.jl software, as in Figure \ref{fig:ODEsSolution}.
\begin{figure}[b]
    \sidecaption
    \includegraphics[width=0.6\columnwidth]{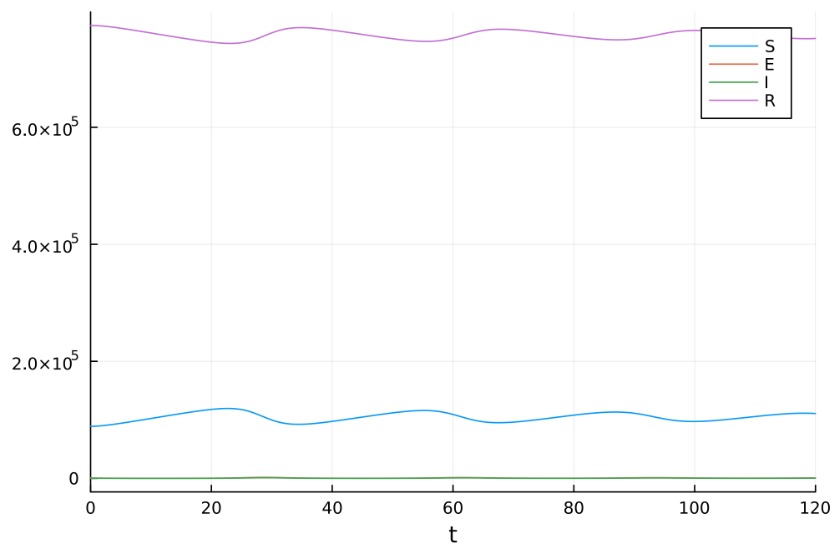}
    \caption{An example solution of the ODEs of the SEIR stock-flow diagram}
    \label{fig:ODEsSolution}
\end{figure}

It is notable that the ODE semantics mapping is more general than illustrated above: we can also map the syntax of ``open'' stock-flow diagrams (to be discussed in Section \ref{section: open stock flow}) to ``open'' dynamical systems.  This supports composition of models.

\subsection{Causal Loop Diagrams}
\label{subsection: causal loop diagrams}

In this section, we define a semantics for stock-flow diagrams that maps any such diagram into a particularly simple kind of causal loop diagram.  In System Dynamics, a ``causal loop diagram'' is a graph where each node represents a variable and each edge represents a way in which one variable can directly influence another \cite{sterman2000business}.  The edges are typically labelled with $\pm$ signs called ``polarities'' that indicate whether an increase in the source variable tends to increase or decrease the target variable, \textit{ceteris paribus}.  From these we can compute signs for loops of edges, which describe positive or negative feedback loops.

Here we consider a simplified variant of causal loop diagram which is simply a graph: the polarities for edges are not included. The problem of interpreting a stock-flow diagram as a fully annotated causal loop diagram is left for future work.

\begin{figure}[h]
    \sidecaption
    \includegraphics[width=0.4\columnwidth]{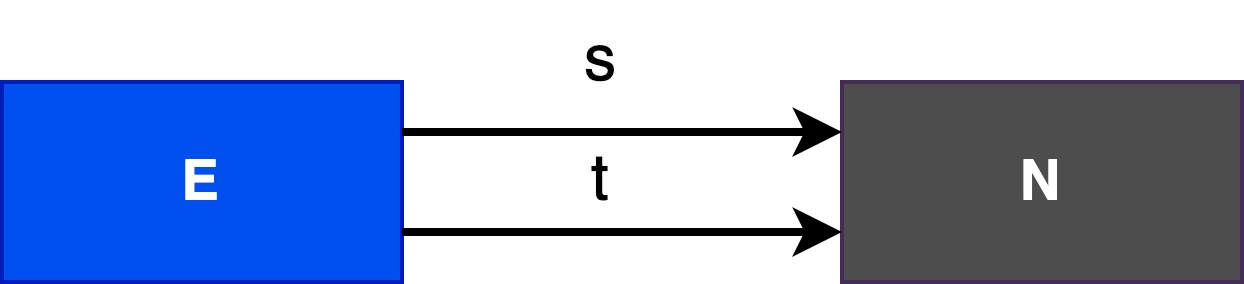}
    \caption{The schema for causal loop diagrams}
    \label{fig:schema_causal_loop}
\end{figure}

Figure \ref{fig:schema_causal_loop} depicts the schema for  causal loop diagrams.  It is just the schema for what category theorists call ``graphs'' (that is, directed multigraphs allowing self-loops).  There is an object $\E$ for edges, an object $\Node$ for nodes, and morphisms $s, t \maps \E \to \Node$ sending each edge to its source and target node, respectively.  Thus, an instance $\CL$ of this schema is simply a finite set $\CL(\Node)$ of nodes, a finite set $\CL(\E)$ of edges, and maps $\CL(s), \CL(t) \maps \CL(\E) \to \CL(\Node)$ sending each edge to its source and target node.

There is in fact a way to translate any stock-flow diagram $(G, \phi)$ into a causal loop diagram $\CL$.   It works as follows:

\begin{enumerate}
  \item{The set of nodes $\CL(\Node)$ is the disjoint union of the set of stocks, the set of sum variables, and the set of auxiliary variables (which includes such a variable for each flow). Explicitly, 
    \[
        \CL(\Node) \coloneqq \G(\S)  \sqcup \G(\SV) \sqcup \G(\V). 
    \]}%The image of the function of $G(\fv)$ is excluded because it repeats the set of flows $G(\F)$.
   \item{The set of edges $\CL(\E)$ is given by
   \[  \CL(\E) := \G(\LV) \sqcup \G(\LS) \sqcup \G(\LSV) \sqcup \G(\I)\sqcup \G(\O) .\]}
  \item{Each edge coming from a variable link \(\ell \in \G(\LV)\) has source \(\G(\lvs)(\ell)\) and target \(\G(\lvv)(\ell)\).  The source and target of edges coming from sum links and sum variable links are defined similarly.  Each edge coming from an inflow \(i \in \G(\I)\) has the auxiliary variable \(\G(\fv) \G(\ifn) (i)\) as its source and the stock \(G(\is) (i)\) as its target. Note that the auxiliary variable \(\G(\fv) \G(\ifn) (i)\) represents the flow for which $i$ is the inflow.  Similarly, each edge coming from an outflow \(o \in \G(\O)\) has the stock \(G(\os)(o)\) as its source and the auxiliary variable \(\G(\fv) \G(\ofn) (o)\) as its target.}
\end{enumerate}

Using this procedure, the SEIR stock-flow diagram in Figure \ref{fig:SEIR_stockflow} gives rise to the causal loop diagram in Figure \ref{fig:seir_causal_loop}.  Here, following tradition, we have labeled some of the nodes by flows rather than their corresponding auxiliary variables, e.g.\ ``birth'' rather than ``vbirth''.  Since the map $\G(\fv) \maps \G(\F) \to \G(\V)$ is usually one-to-one, this does not create ambiguities in practice.

\begin{figure}[H]
    \sidecaption
    \includegraphics[width=0.6\columnwidth]{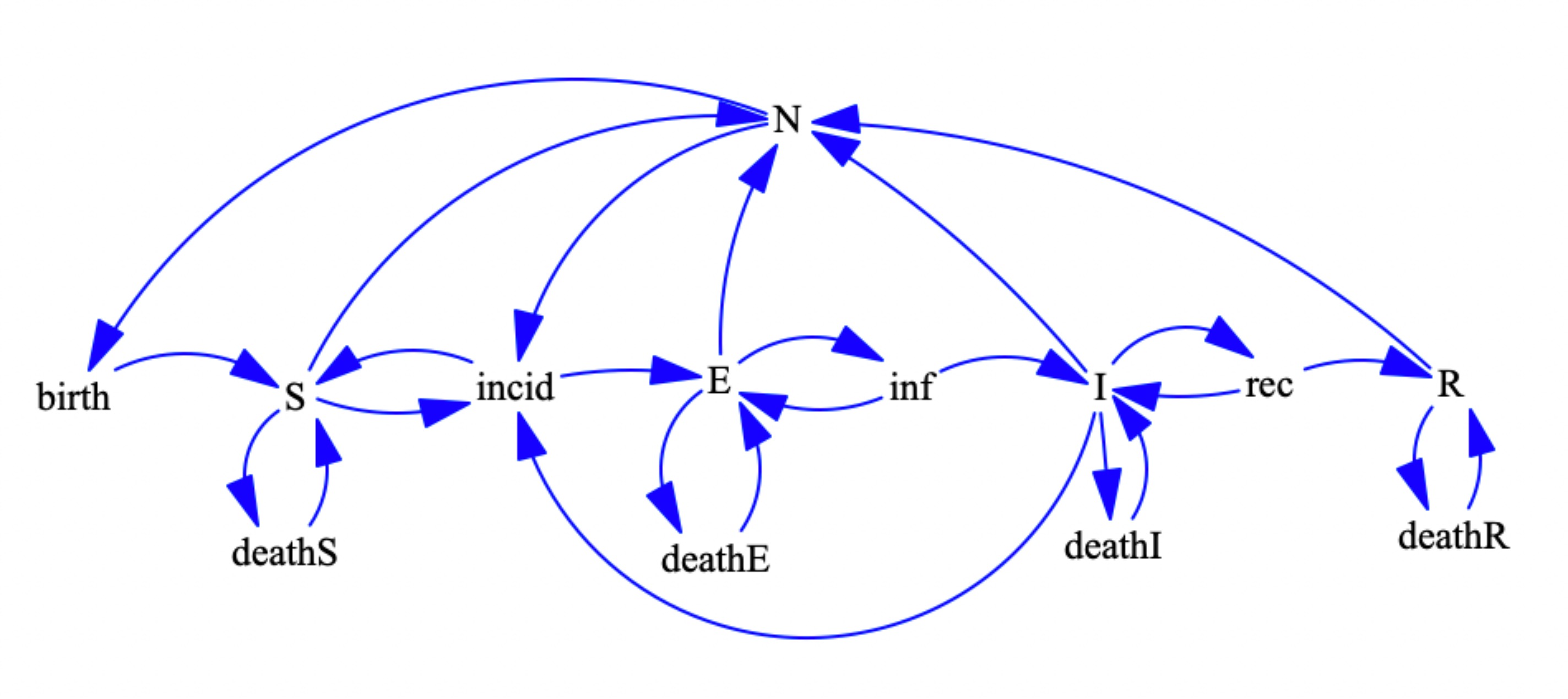}
    \caption{The causal loop diagram of the SEIR model}
    \label{fig:seir_causal_loop}
\end{figure}

Just as with stock-flow diagrams, we use the data structure of a categorical database to encode causal loop diagrams in the software. Figure \ref{fig:schema_database_causal_loop} shows an example: the categorical database for the SEIR causal loop diagram.

\begin{figure}[h]
    \sidecaption
    \includegraphics[width=0.6\columnwidth]{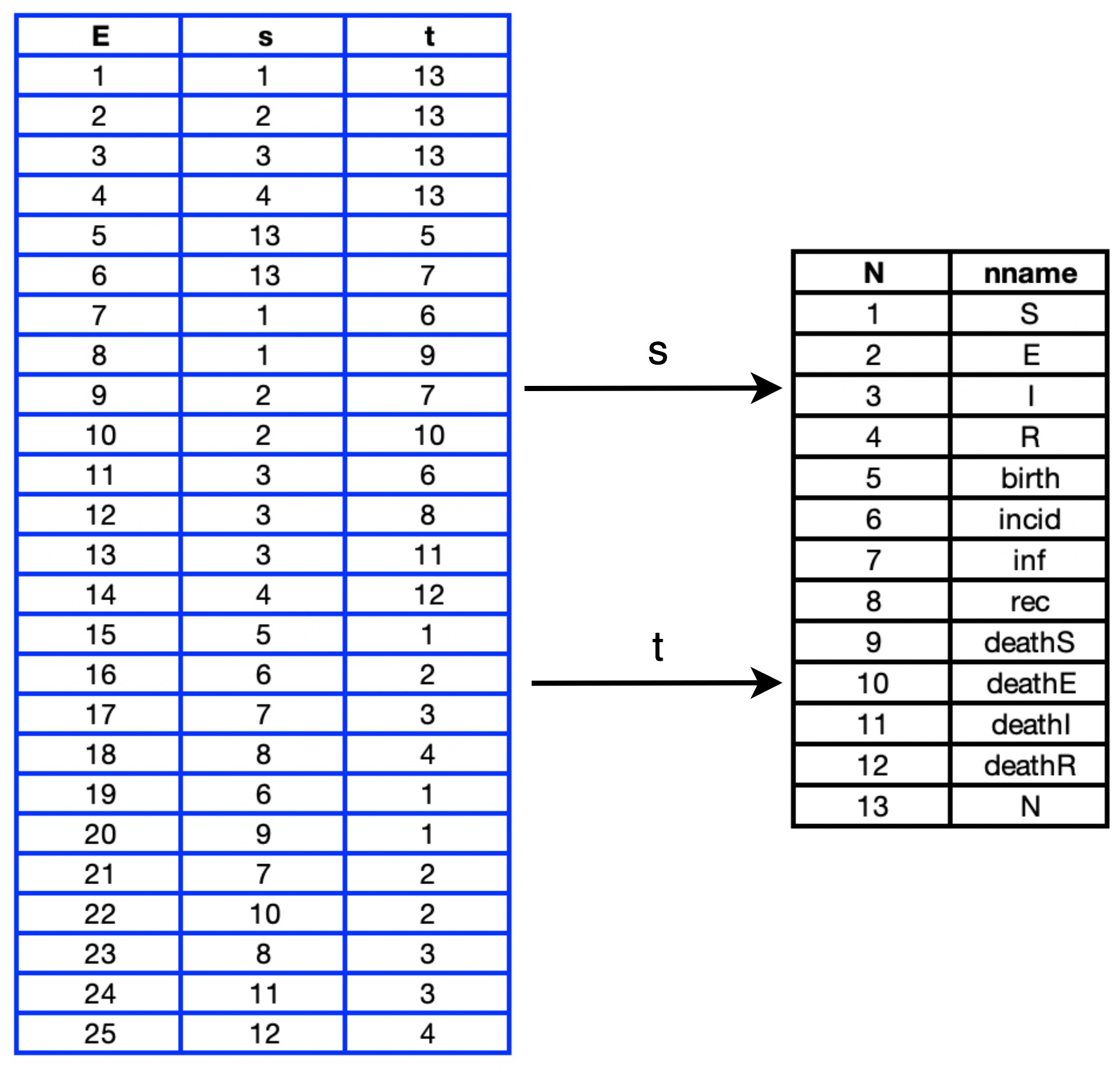}
    \caption{The categorical database structure representing the SEIR causal loop diagram}
    \label{fig:schema_database_causal_loop}
\end{figure}

\subsection{System Structure Diagrams}
\label{subsection: system structure}

In System Dynamics \cite{sterman2000business}, a system structure diagram is a \emph{purely qualitative} version of a stock-flow diagram.  It lacks the quantitative information provided by the functions $\phi_v$ that describe how each auxiliary variable $v$ depends on the quantities linked to it. While many uses of system structure diagrams further annotate links in such diagrams with polarities, we reserve for future work the problem of automatic derivation of such polarities.  For now, we therefore define a ``system structure diagram'' to be simply an instance $G$ of the schema for stock-flow diagrams, as explained in Section \ref{section: closed stock flow}.  

There is a semantics for stock-flow diagrams that maps them to system structure diagrams: it simply maps any stock-flow diagram $(G,\phi)$ to the system structure diagram $G$.   In fact one can construct a category $\StockFlow$ of stock-flow diagrams, a category $\SystemStructure$ of system structure diagrams, and a functor
\[
\begin{tikzpicture}[baseline=(current  bounding  box.center),scale=1.5]
\node (A) at (0,0) {$\StockFlow$};
\node (B) at (2,0) 
{$\SystemStructure$};
\path[->,font=\scriptsize]
(A) edge node [above] {$A$} (B);
\end{tikzpicture}
\]
that maps any stock-flow diagram $(G,\phi)$ to the system structure diagram $G$.  

Likewise, the semantics described in Section \ref{subsection: causal loop diagrams} gives a functor 
\[
\begin{tikzpicture}[baseline=(current  bounding  box.center),scale=1.5]
\node (A) at (0,0) {$\StockFlow$};
\node (B) at (2,0) 
{$\CausalLoop$};
\path[->,font=\scriptsize]
(A) edge node [above] {$B$} (B);
\end{tikzpicture}
\]
from the category of stock-flow diagrams to a suitable category $\CausalLoop$.  But note that the causal loop diagram associated to a stock-flow diagram $(G,\phi)$ depends only on $G$: the functions $\phi_v$ play no role here, though they would for the more elaborate and more commonly used causal loop diagrams with edges labeled by polarities.  Thus, our causal loop and system structure semantics for stock-flow diagrams fit into a so-called ``commutative diagram'' of functors between categories:
\[
\begin{tikzpicture}[baseline=(current  bounding  box.center),scale=1.5]
\node (A) at (0,0) {$\StockFlow$};
\node (B) at (1,-1) 
{$\SystemStructure$};
\node (C) at (2,0) 
{$\CausalLoop$};
\path[->,font=\scriptsize]
(A) edge node [left] {$A$} (B)
(A) edge node [above] {$B$} (C)
(B) edge node [right] {$C$} (C);
\end{tikzpicture}
\]
In essence, this diagram says that to turn a stock-flow diagram $(G,\phi)$ into a causal loop diagram in the manner explained in Section \ref{subsection: causal loop diagrams}, we can first extract the system structure diagram $G$ and then turn that into a causal loop diagram.

This commutative diagram illustrates a general fact: rather than the various semantics for a given form of syntax being separate from each other, like isolated walled gardens, they are often related in fruitful ways.  Category theory lets us formalize these relationships, and category-based software lets us apply them in practical ways. 

\section{Composing Open Stock-Flow Diagrams}
\label{section: open stock flow}

We can build larger stock-flow diagrams by gluing together smaller ones.  To achieve this goal, we need ``open'' stock-flow diagrams.  An example is shown in Figure \ref{fig:structured_multiple_cospan}.  It consists of a stock-flow diagram together with some extra data describing interfaces at which we can compose this diagram to other stock-flow diagrams.

\begin{figure}[H]
    \centering
    \includegraphics[width=0.8\columnwidth]{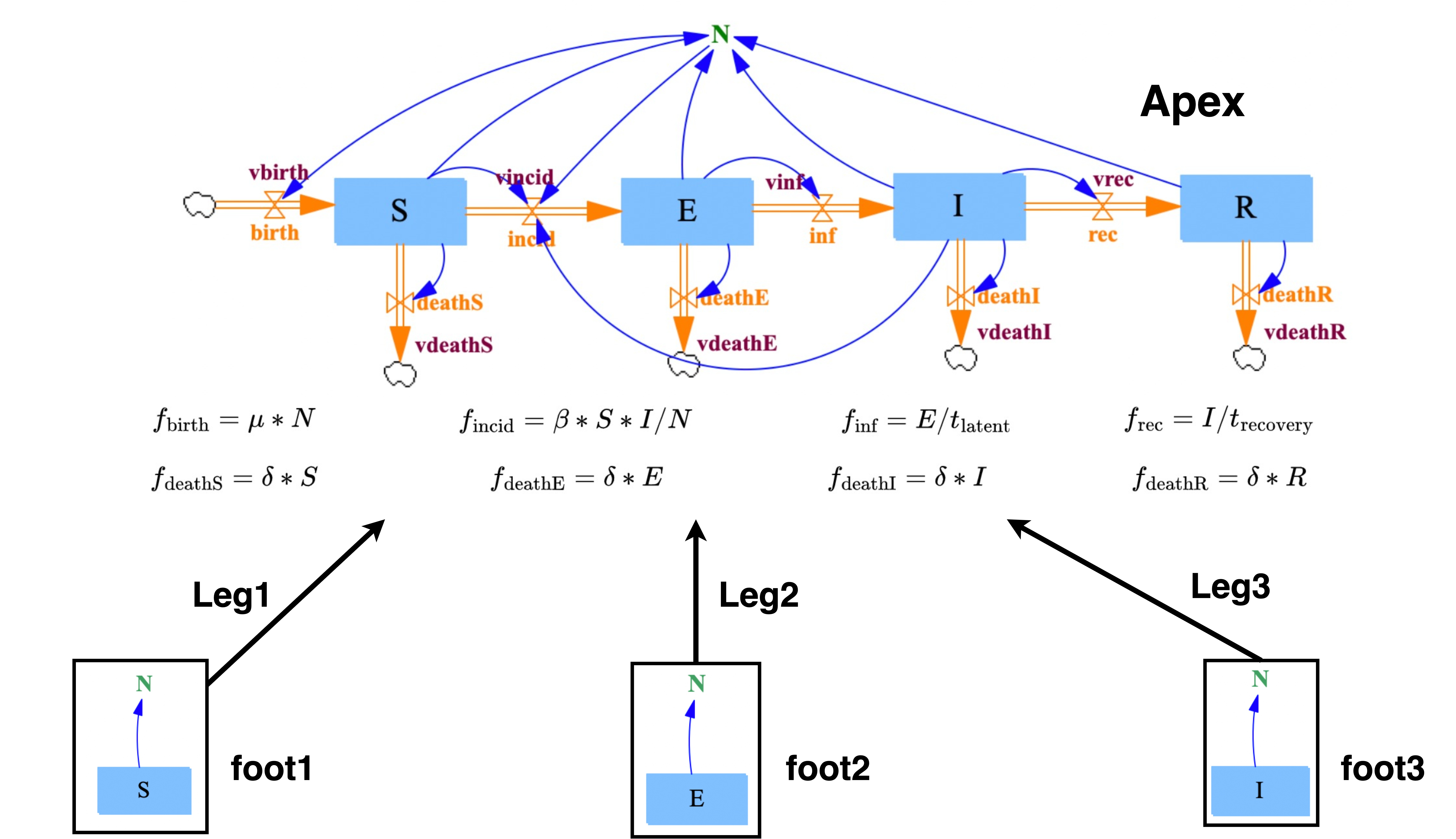}
    \caption{An example of an open stock-flow diagram}
    \label{fig:structured_multiple_cospan}
\end{figure}

More precisely, an open stock-flow diagram consists of an ``apex'' together with a finite collection of ``feet'' and ``legs''.   The ``apex'' is any stock-flow diagram $(G,\phi)$.  Each ``foot'' is also a stock-flow diagram, and it comes with a ``leg'', which is a map of stock-flow diagrams from the foot to $(G,\phi)$.   However, we require that the feet are stock-flow diagrams of a restricted sort: they can contain only stocks, sum variables and sum links.  We enforce this restriction because we want to glue together stock-flow diagrams only by identifying certain stocks in one diagram with stocks in another diagram; however, doing this properly may require identifying certain sum variables and sum links as well. 

\begin{figure}[H]   \sidecaption
    \includegraphics[width=0.35\columnwidth]{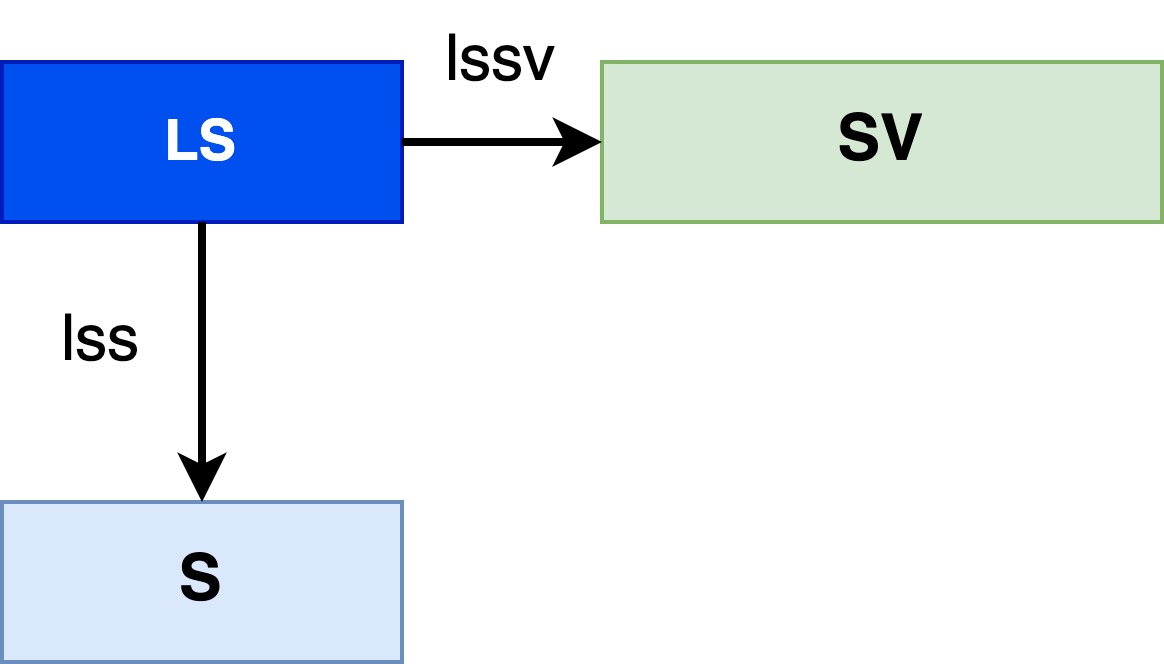}
    \caption{The schema for interfaces}
    \label{fig:footschema}
\end{figure}

We call these restricted stock-flow diagrams ``interfaces''.   Just as there is a schema for stock-flow diagrams, there is a schema for interfaces, shown in Figure \ref{fig:footschema}.  An interface, say $X$, is precisely an instance of this schema. It thus consists of:
\begin{enumerate}
\item a finite set $X(\S)$ of stocks,
\item a finite set of sum variables $X(\SV)$, a finite set of sum links $X(\LS)$, and functions
\[  X(\lss) \maps X(\LS) \to X(\S), \qquad
X(\lssv) \maps X(\LS) \to X(\SV) .\]
\end{enumerate}

In our previous paper \cite{baez2022compositional} we explain how to compose open stock-flow diagrams.  Briefly, we use the mathematics of ``decorated cospans'' \cite{fong2015}---a general category-theoretic framework for the composition of open systems. However, to implement this in code in StockFlow.jl we also take advantage of a closely related framework, ``structured cospans'' \cite{baezcourser2020,baez-courser-vasilakopoulou2022}. 
The reason is that structured cospans have already been systematically implemented in the Catlab.jl package \cite{patterson-lynch-fairbanks2021}, which serves as the foundation underlying our implementation of StockFlow.jl.  StockFlow.jl is one of the newer members of the AlgebraicJulia ecosystem   \cite{AlgebraicJulia}, which provides computational support for applied category theory via Catlab.jl.

Two other twists are also worth noting, if only for cognoscenti.  Firstly, in  StockFlow.jl we actually use decorated or structured ``multicospans'' \cite{libkind-baas-patterson-fairbanks2021,spivak2013} instead of cospans.  The difference is that multicospans can have multiple legs and feet, as shown for example in Figure \ref{fig:structured_multiple_cospan}, while cospans have exactly two. Secondly, StockFlow.jl applies the flexible graphical syntax of undirected wiring diagrams \cite{fongspivak2018} to compose these multicospans.

However, users do not need to understand these technicalities to use our software.  The key ideas can be understood from an example.  Figure \ref{fig:structured_multiple_cospan} shows an open stock-flow diagram with three legs and feet: this is an SEIR model with three interfaces containing the stocks $\S, \E$ and $\I$, respectively.   Figure \ref{fig:composition} shows an example of composition in which this open stock-flow diagram is glued to another open stock-flow diagram, an SVEI model, along all three interfaces.  The result of composition is yet another open stock-flow diagram, but in this particular case there are no interfaces left, so it amounts to an ordinary stock-flow diagram.  The composite diagram describes an SEIRV model.  Like any other diagram, diagrams that result from such composition can be mapped to alternative semantic domains. For example, the bottom right plot in Figure \ref{fig:composition} then shows a solution of the system of ODEs to which the composite stock-flow diagram is mapped using ODE semantics.

\begin{figure}[H]
    \centering
    \includegraphics[width=1\columnwidth]{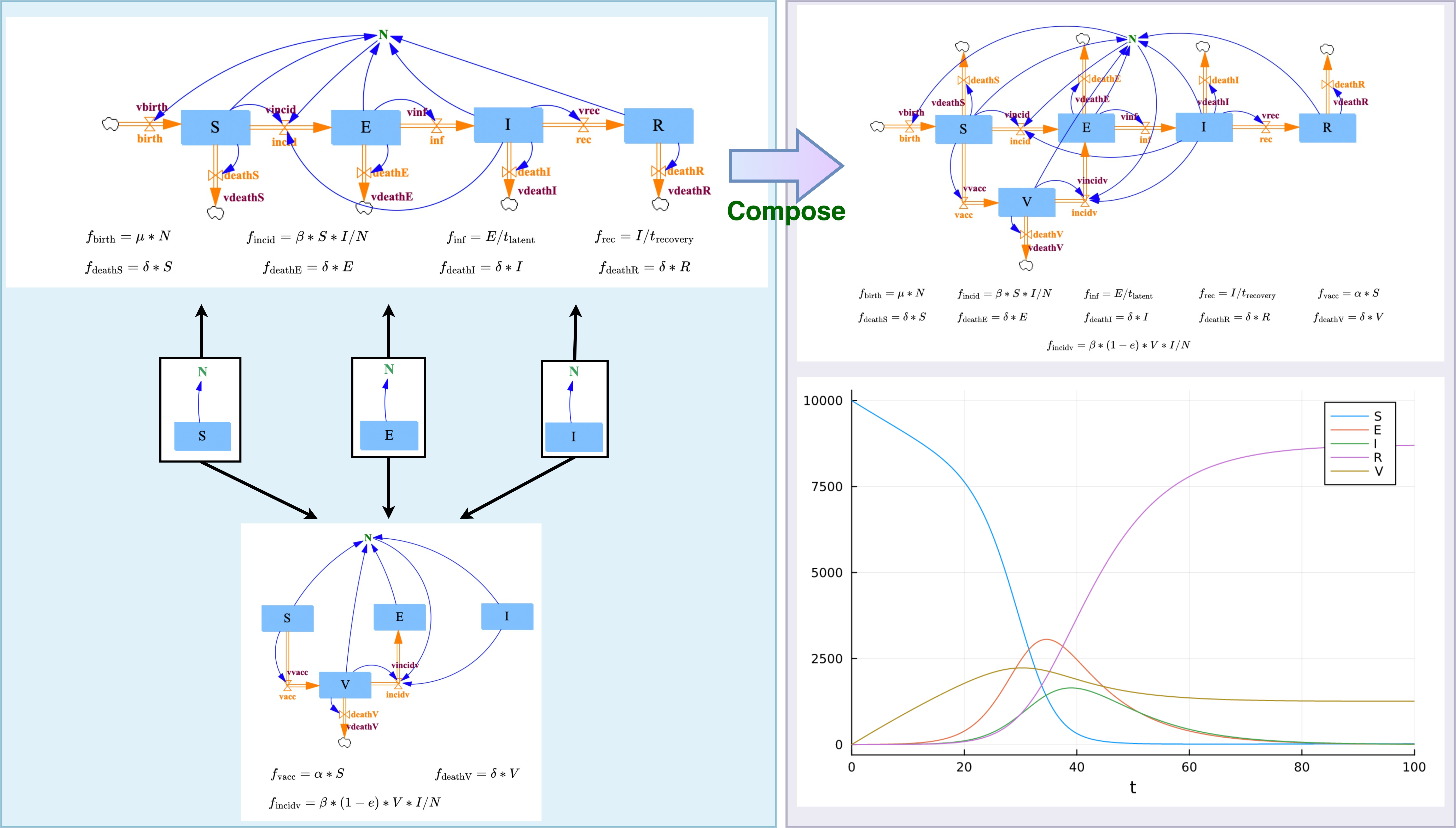}
    \caption{Composing SEIR and SVEI models to obtain an SEIRV model}
    \label{fig:composition}
\end{figure}

\section{Stratifying Typed System Structure Diagrams}
\label{section: stratification}

Just as we introduced \emph{open} stock-flow diagrams to permit \emph{composition} of models, we now introduce \emph{typed} diagrams to support \emph{stratification} of models.  Recall from Section \ref{section: introduction} that ``stratifying'' a model involves breaking stocks into smaller stocks that differ in some characteristics.  For example, we might take the simple SEIR model shown in Figure \ref{fig:SEIR_stockflow} and subdivide each stock into two groups of different sexes, or three groups of different ages, or both.  This subdivision is a common and important procedure for refining models in epidemiology.

However, stratification also requires introducing new flows, new auxiliary variables, new links, and so on.  For example, consider the flow ``$\textrm{inf}$'' from the ``Exposed'' stock to the ``Infectious'' stock in the SEIR model.  If we stratify this model by breaking each stock into two sexes, we also need to replace this flow with two separate flows, one for each sex.  Furthermore, since the rate of this flow is given by an auxiliary variable, we must replace that variable by two separate variables if we wish to allow the two sexes to have different rates of infection---which indeed is the whole point of stratifying the model in this way.

The challenge is to carry out all these steps in a mathematically well-defined way that can be cleanly implemented in software.  Libkind \textit{et al.} recently did this for a related diagrammatic modeling language that has also been implemented in the AlgebraicJulia ecosystem: Petri nets \cite{libkind2022epidemic}.  The key was to introduce ``typed'' Petri nets and use pullbacks, a standard construction in category theory \cite{leinster}.  However, their approach to stratification has the potential to be generalized to many other diagrammatic modeling languages.  Here we adapt their approach to stock-flow diagrams. 

To begin, it is important to realize that when we stratify a model described by a stock-flow diagram \((G,\phi)\), we do \emph{not} expect that all the functions \(\phi_v\) associated to auxiliary variables \(v\) can be copied over from the original model in an automatic way.   Indeed, the point of stratification is to let these functions depend on the ``stratum'', e.g.\ the sex, the age group, and so on.  Thus, in the approach we take here, we first stratify not the whole stock-flow diagram \((G,\phi)\) but only its underlying system structure diagram \(G\) (as defined in Section \ref{subsection: system structure}). To promote the stratified system structure diagram to a stock-flow diagram, the user must then choose functions for the auxiliary variables.  In future work, we can enable an approach where most, but not all, of the original functions are automatically reused.

How do we stratify system structure diagrams?  A first naive thought would be to take a ``product'' of two system structure diagrams:
\begin{enumerate}
\item
The original system structure diagram that we wish to stratify.  We call this the ``aggregate model'' and denote it as $S_\tx{aggregate}$.   
\item
A system structure diagram describing the strata (e.g., age groups or sexes) that we wish to use in stratifying the aggregate model.  We call this the ``strata model'' and denote it as $S_\tx{strata}$.  
\end{enumerate}
The product of these two, denoted \(S_\tx{aggregate} \times S_\tx{strata}\), is a system structure diagram for which a stock is an ordered pair \((x,y)\) consisting of a stock \(x\) in the aggregate model \(S_\tx{aggregate}\) and a stock \(y\) in the strata model \(S_\tx{strata}\). Similarly a flow in the product is an ordered pair of flows, one from each model---and so on for inflows, outflows, auxiliary variables, and all the other objects in the schema for stock-flow diagrams.

In fact, category theory has a general notion of ``product'' \cite{leinster} applicable to any category, and  \(S_\tx{aggregate} \times S_\tx{strata}\) is a product in the category \(\SystemStructure\).  One consequence is that it comes with maps as follows:
\[
\begin{tikzpicture}[baseline=(current  bounding  box.center),scale=2]
\node (A) at (0,1) {$S_\tx{aggregate} \times S_\tx{strata}$};
\node (B) at (1.5,1) {$S_\tx{aggregate}$};
\node (C) at (0,0) {$S_\tx{strata}$};
\path[->,font=\scriptsize]
(A) edge node [above] {$p_1$} (B)
(A) edge node [left] {$p_2$} (C);
\end{tikzpicture}
\]

Given any ordered pair \((x,y)\) of stocks, flows, etc.\ in the product \(S_\tx{aggregate} \times S_\tx{strata}\), the maps \(p_1\) and \(p_2\) pick out the components of this ordered pair:
\[   p_1(x,y) = x, \qquad p_2(x,y) = y .\]

Unfortunately, the product \(S_\tx{aggregate} \times S_\tx{strata}\) often contains more stocks, flows, etc.\ than we really want.  The solution is to keep only the ordered pairs \((x,y)\) where \(x\) and \(y\) have the same ``type''.   To do this, we introduce a third stock-flow diagram \(S_\tx{type}\), called the ``type system'', together with maps 
\[
\begin{tikzpicture}[baseline=(current  bounding  box.center),scale=2]
\node (A) at (0,1) {};
\node (B) at (1.5,1) {$S_\tx{aggregate}$};
\node (C) at (0,0) {$S_\tx{strata}$};
\node (D) at (1.5,0) {$S_\tx{type}$};
\path[->,font=\scriptsize]
(B) edge node [right] {$t_\tx{aggregate}$} (D)
(C) edge node [below] {$t_\tx{strata}$} (D);
\end{tikzpicture}
\]

We obtain the desired stratified model $S_\tx{stratified}$ by taking a ``pullback'' in the category $\StockFlow$.  The pullback is a particular stock-flow diagram equipped with maps making the following square commute:
\[
\begin{tikzpicture}[baseline=(current  bounding  box.center),scale=2]
\node (A) at (0,1) {$S_\tx{stratified}$};
\node (B) at (1.5,1) {$S_\tx{aggregate}$};
\node (C) at (0,0) {$S_\tx{strata}$};
\node (D) at (1.5,0) {$S_\tx{type}$};
\path[->,font=\scriptsize]
(A) edge node [above] {$p_1$} (B)
(A) edge node [left] {$p_2$} (C)
(B) edge node [right] {$t_\tx{aggregate}$} (D)
(C) edge node [below] {$t_\tx{strata}$} (D);
\end{tikzpicture}
\]
The pullback is defined so that a stock in $S_\tx{stratified}$ is an ordered pair \((x,y)\) consisting of a stock $x$ in the aggregate model $S_\tx{aggregate}$ and a stock $y$ in the strata model $S_\tx{strata}$ that both map to the same stock in the type system $S_\tx{type}$. In other words, the pair $(x,y)$ satisfies
\[    (t_\tx{aggregate})_\S (x) = (t_\tx{strata})_\S(y) .\]
Similarly, a flow in $S_\tx{stratified}$ is a pair consisting of a flow in the aggregate model and a flow in the strata model that both map to the same flow in the type system---and so on for inflows, outflows, auxiliary variables, and all the other objects in the schema for stock-flow diagrams.  As before, the maps \(p_1\) and \(p_2\) pick out the components of these ordered pairs:
\[   p_1(x,y) = x, \qquad p_2(x,y) = y .\]
All this and more follows from the general theory of pullbacks, which works in any category \cite{leinster}.  That is why this approach to stratification works so generally.

Before we turn to examples of stratification, let us briefly explain the maps between system structure diagrams that appear as arrows in the above diagram. Technically these maps are the \emph{morphisms} in the category $\SystemStructure$, and they play an important role in the theory.  But what are these maps like?  

Given system structure diagrams \(\G\) and \(\H\), a map $\alpha \maps \G \to \H$ consists of functions sending all the stocks, flows, inflows, outflows, auxiliary variables, etc.\ for $\G$ to the corresponding items for $\H$. We use $\alpha_\S \maps \G(\S) \to \H(\S)$ to denote the function on stocks, $\alpha_\F \maps \G(\F) \to \H(\F)$ to denote the function on flows, and so forth.  But we require that these functions be structure-preserving.   For example, if \(\alpha_\F\) maps a flow \(f \in \G(\F)\) to a flow \(f' \in \H(\F)\), then we require that \(\alpha_\F\) must map the upstream of \(f\) to the upstream of \(f'\), and map the downstream of \(f\) to the downstream of \(f'\). For details, see \cite{baez2022compositional}.

Figure \ref{fig:type_morphism_example} shows an example of such a map $\alpha$ from the SEIR system structure diagram (top) to the SIR system structure diagram (bottom). To show the map clearly, we have coloured the components of the SEIR diagram according to the colour of the component in the SIR diagram to which they are mapped.  Notice that $\alpha$ maps the flow $\tx{inf}_2$ of the SEIR diagram to the flow $\tx{inf}_1$ of the SIR diagram.  Thus, we require that $\alpha$ maps the upstream stock of $\tx{inf}_2$ to the upstream stock of $\tx{inf}_1$, and the downstream stock of $\tx{inf}_2$ to the downstream stock of $\tx{inf}_1$. 

\iffalse
Explicitly, $\alpha_S(S_2) = S_1$ and $\alpha_S(E_2) = I_1$, and Figure~\ref{fig:type_morphism_example} specifies that $S_2$ is the upstream stock of $\tx{inf}_2$, $S_1$ is the upstream stock of $\tx{inf}_1$, $E_2$ is the downstream stock of $\tx{inf}_2$, and $I_1$ is the downstream stock of $\tx{inf}_1$.  The maps for other parts of the system structure diagram must be similarly coordinated.
\fi

\begin{figure}[H] \sidecaption
    \centering
    \includegraphics[width=0.6\columnwidth]{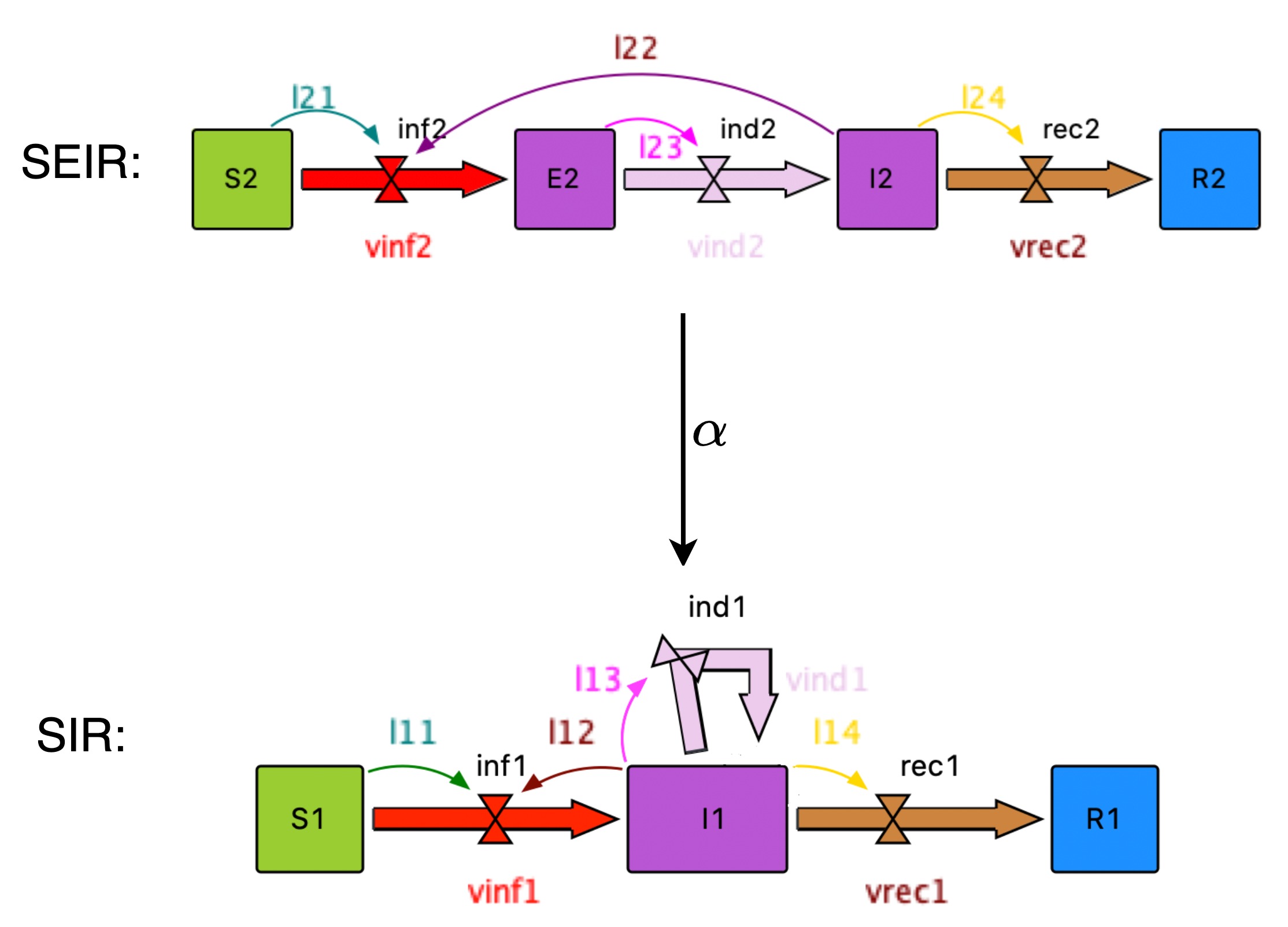}
    \caption{An example of a map (morphism) from the SEIR system structure diagram to the SIR system structure diagram}
    \label{fig:type_morphism_example}
\end{figure}

\begin{figure}[b]
    \sidecaption
    \includegraphics[width=0.5\columnwidth]{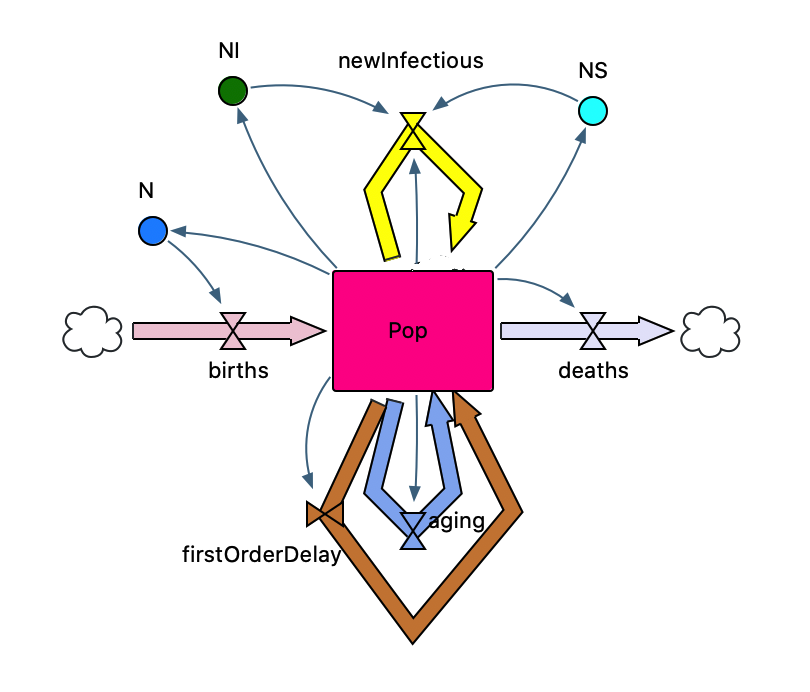}
    \caption{A stock-flow diagram $S_\tx{type}$ serving as a type system for infectious disease models}
    \label{fig:s_type}
\end{figure}

Now let us turn to examples of stratification.   Figure \ref{fig:s_type} shows a system structure diagram \(S_\tx{type}\) that can serve as a type system for stratified infectious disease models.  This system structure diagram has just one stock, \(\textrm{Pop}\), which represents all the kinds of populations.  It has five flows, representing five different types of flows in the infectious disease models we wish to build: 
    \begin{enumerate}
        \item{the birth flow,}
        \item{the death flow,}
        \item{the flow for new (incident) infections,}
        \item{the aging flow representing the transition from one age group to its immediately older group,}
        \item{the first order delay flow based on the schema of the system structure diagrams.}
    \end{enumerate}
To support a clearer visualization, these flows are drawn using five different colours.   The type system \(S_\tx{type}\) has five auxiliary variables corresponding to these five flows---but for simplicity, we do not depict these auxiliary variables.  It also has three sum auxiliary variables: 
    \begin{enumerate}
        \item{\(\N\), representing the total population of the whole model,}
        \item{\(\NS\), representing the population of a specific subgroup,}
        \item{\(\NI\), representing the count of infectious persons of a specific subgroup.}
    \end{enumerate}
Finally, the type system \(S_\tx{type}\) has nine links. 

Figure \ref{fig:examples_typed} shows four system structure diagrams ``typed'' by \(S_\tx{type}\): that is, equipped with maps to \(S_\tx{type}\).  The first two are infectious disease models which can serve as aggregate models $S_\tx{aggregate}$: an SEIR model and an SIS (Susceptible--Infectious--Susceptible) model.  The second two can serve as strata models $S_\tx{strata}$: a sex strata model and an age strata model.   The ``typing'' of these four stock-flow diagrams---that is, their maps to $S_\tx{type}$---are indicated by colours following the colouring scheme of Figure \ref{fig:s_type}.

\begin{figure}
     \centering
     \begin{subfigure}[b]{0.49\textwidth}
         \centering
         \includegraphics[width=\textwidth]{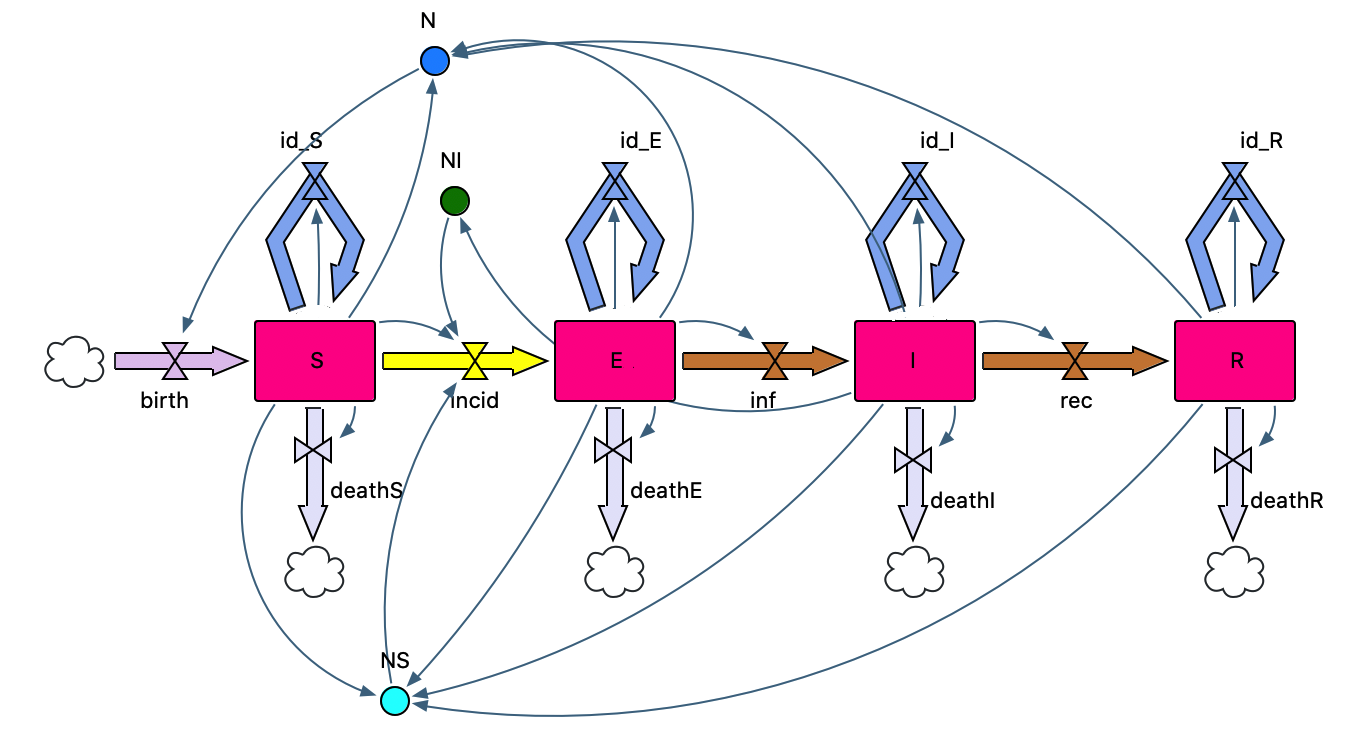}
         \caption{Typed SEIR disease model}
         \label{fig:Typed SEIR disease model}
     \end{subfigure}
     \hfill
     \begin{subfigure}[b]{0.49\textwidth}
         \centering
         \includegraphics[width=\textwidth]{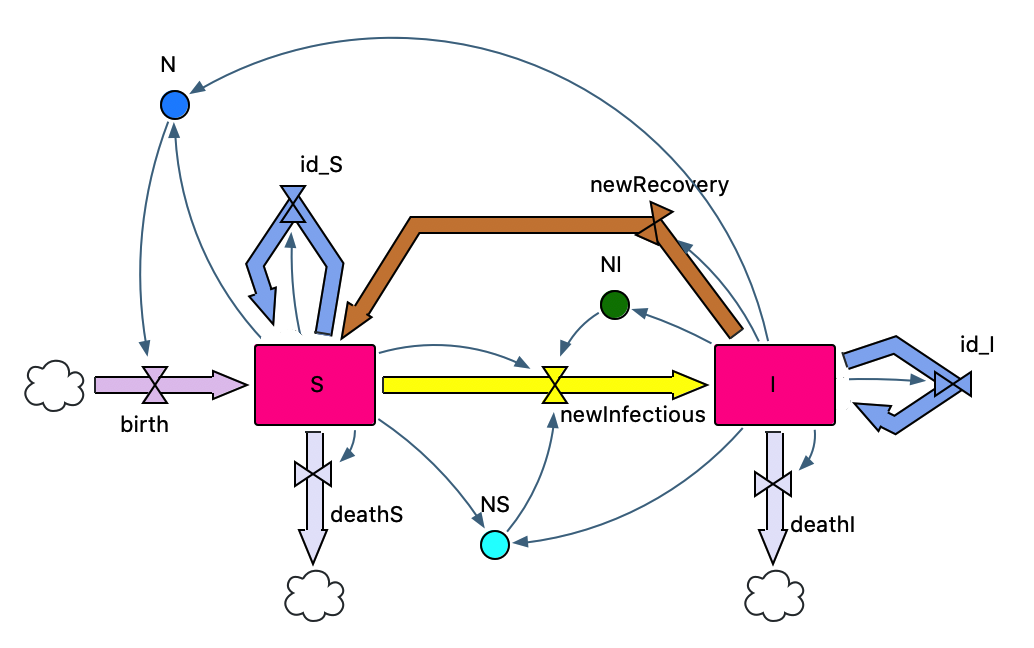}
         \caption{Typed SIS disease model}
         \label{fig:Typed SIS disease model}
     \end{subfigure}
     \hfill
     \begin{subfigure}[b]{0.49\textwidth}
         \centering
         \includegraphics[width=\textwidth]{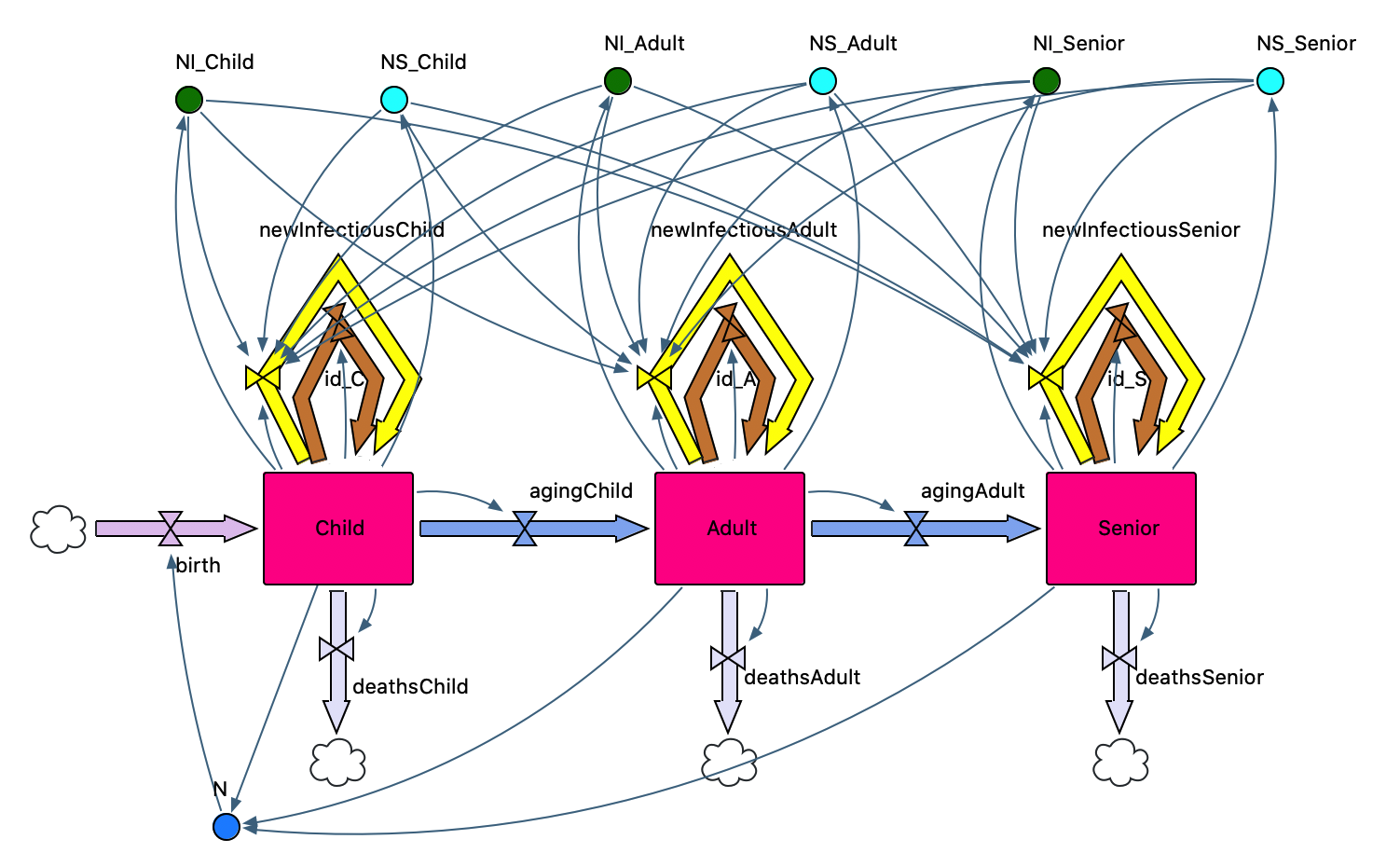}
         \caption{Typed age strata model}
         \label{fig:Typed age strata model}
     \end{subfigure}
     \hfill
     \begin{subfigure}[b]{0.4\textwidth}
         \centering
         \includegraphics[width=\textwidth]{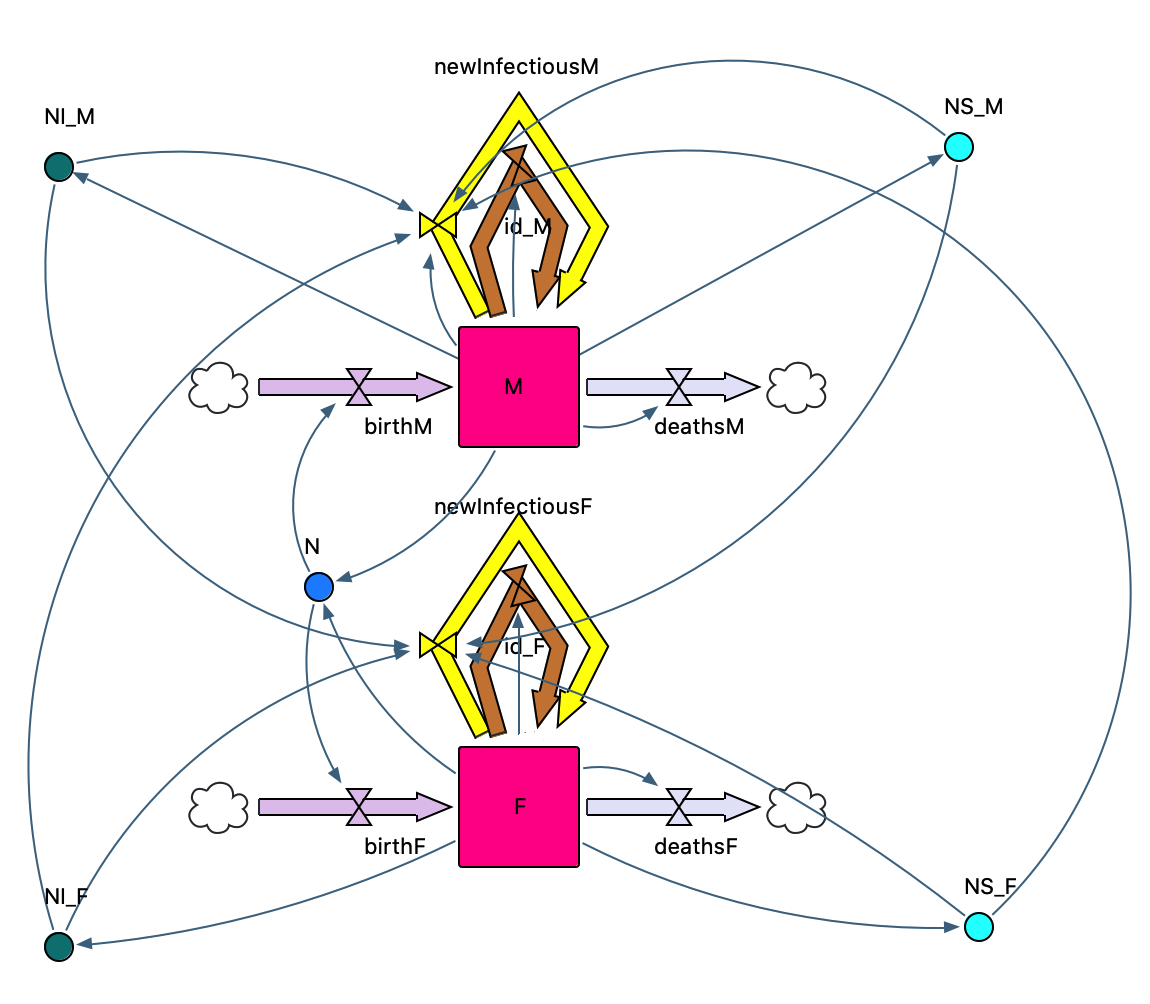}
         \caption{Typed sex strata model}
         \label{fig:Typed sex strata model}
     \end{subfigure}
        \caption{Four examples of typed stock-flow diagrams}
        \label{fig:examples_typed}
\end{figure}

In Figure \ref{fig:stratifications} we show the results of pullback-based stratification for each combination of an aggregate model (either the SEIR model or SIS model) and a strata model (either the age strata model and sex strata model).  A stratified model \(S_\tx{stratified}\) is generated by taking the pullback of each of the four combinations of an aggregate model and a strata model.  These four stratified models are the age-stratified SEIR model, age-stratified SIS model, sex-stratified SEIR model and sex-stratified SIS model. 

More generally, we can define many other system structure diagrams \(S_\tx{aggregate}\) to serve as aggregate models for infectious disease \cite{anderson1992infectious}.  Similarly, we can define many different system structure diagrams \(S_\tx{strata}\) to serve as strata models characterizing the structure and patterns of progression associated with different types of stratification: not only sex and age but also by socioeconomic or employment status, and geographical stratification including mobility amongst regions, mobility amongst regions from a home base in a particular region, etc.   Once an aggregate model and strata model are chosen along with their typings \(t_\tx{aggregate} \maps S_\tx{aggregate} \to S_\tx{type}\) and \(t_\tx{strata} \maps S_\tx{strata}\to S_\tx{type}\), 
our code can automatically construct a stratified model by taking a pullback.   For example, we can generate an SEIR age-stratified system structure diagram by calculating the pullback of the SEIR model and an age strata model. 

\begin{figure}[H]
    \centering
    \includegraphics[width=1\columnwidth]{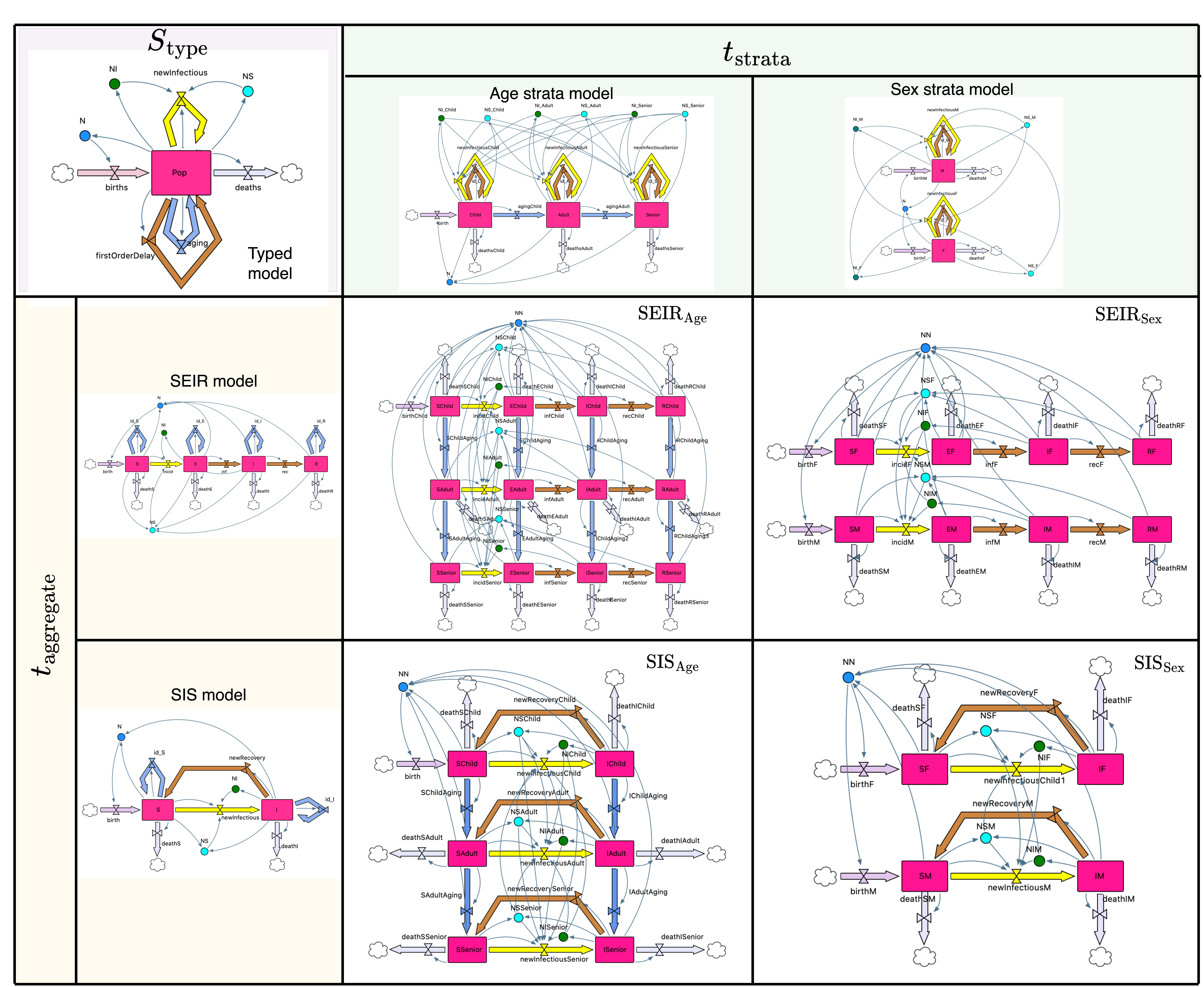}
    \caption{Examples of four stratified models}
    \label{fig:stratifications}
\end{figure}

Moreover, we can build stratified models with multiple dimensions by taking repeated pullbacks of multiple stock-flow diagrams. For example, we can build an age-and-sex stratified SEIR model by such an iterated pullback involving strata models for each of two dimensions---sex and age.  The result is shown in Figure \ref{fig:stratifications_multiple}.  Similar approaches can be used to model progression of multiple comorbidities and behavioural risk factors---a form of stratification that, done by hand, would be subject to a combinatorial explosion of detail  \cite{osgood2009Comorbidities}.

As mentioned, the stratified models here are built based on system structure diagrams, with the goal of simplifying the stratification process and avoiding the need to consider the functions in the stock-flow diagrams. Fully stratified stock-flow diagrams are then generated by assigning functions to each auxiliary variable of the stratified system structure diagram. Figure \ref{fig:solution_stratified_sis_sex} shows a solution of an example SIS model stratified by sex.

\begin{figure}[h]
    \centering
    \includegraphics[width=1.0\columnwidth]{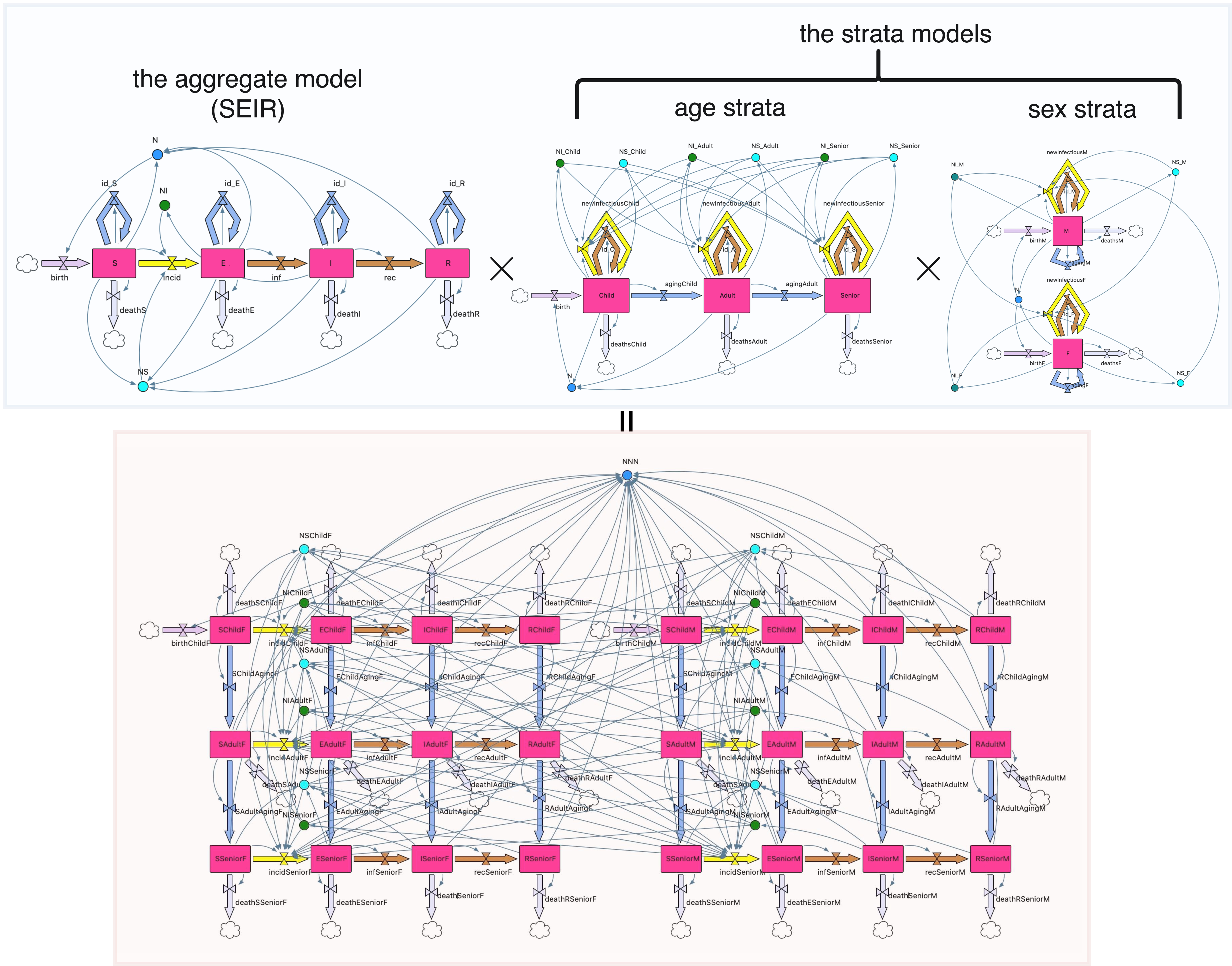}
    \caption{Example of building an SEIR model stratified in multiple dimensions}
    \label{fig:stratifications_multiple}
\end{figure}

\begin{figure}[h]
    \sidecaption
    \includegraphics[width=0.6\columnwidth]{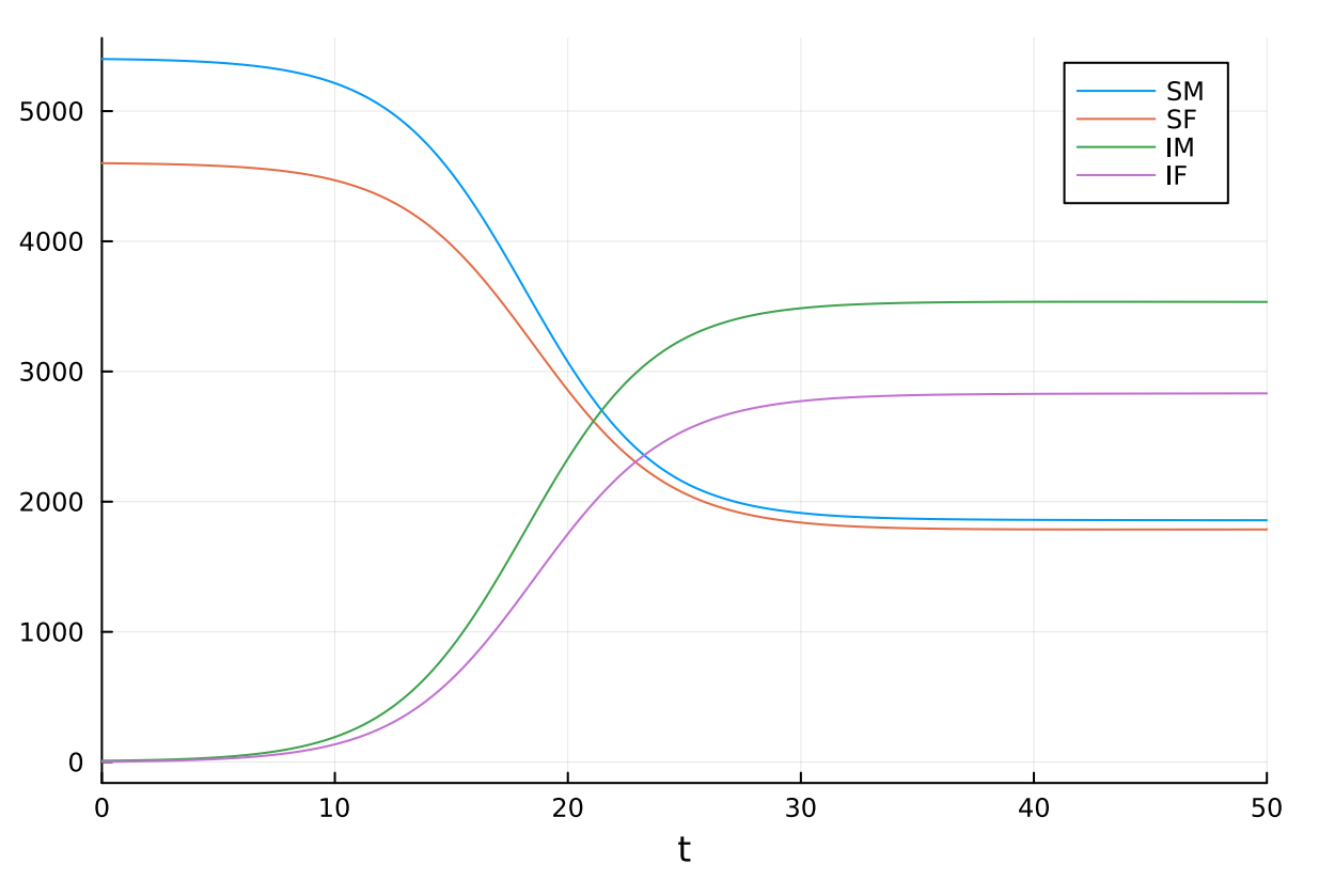}
    \caption{Solutions of the ODE semantics for a sex-stratified SIS model}
    \label{fig:solution_stratified_sis_sex}
\end{figure}

\section{ModelCollab: A Graphical Real-time Collaborative Compositional Modeling Tool}
\label{section: ModelCollab}

While the modular modeling approaches explored in this chapter make ubiquitous use of category theory, use of the resulting functionality---for example, the ability to compose and stratify diagrams, or to interpret them through differing analyses---does not require knowledge of that foundation. Impactful infectious disease modeling is typically conducted in interdisciplinary teams, and securing timely, ongoing feedback about model structure and emergent behaviour from non-modeler team members is key to both model refinement and organizational learning from modeling. Often tacit knowledge of non-modeling team members concerning the system under study (say, evidence for episodic reemergence of a communicable disease in certain demographic segments despite prevention and control efforts) is only elicited once team members have a chance to comment on visualizations of model structure and summaries of model dynamics. Often interpretation of such dynamics is greatly enhanced through reasoning about the relationship between observed behaviour and the diagram structure -- for example, through recognizing that an increase in a variable over time reflects a situation where inflow is greater than outflow, explaining an invariant value of a state variable in terms of a balance between inflows and outflows, or reasoning about exponential change in terms of driving feedback loops. 

Partly for these reasons, the System Dynamics tradition of modeling has long prized the use of visual modeling software that keeps the attention of modelers -- and other team members -- on diagrams depicting model structure. While such software does support communication across interdisciplinary team members, it sufferings both from the disadvantages of traditional treatment of such models discussed in the introduction and a limit to being modified---and often viewed---by only a single user at a time. 

We describe here open-source, visual, collaborative, categorically-rooted, diagram-centric software for building, manipulating, composing, and analyzing System Dynamics models.  This web-based software, named ModelCollab, is designed for real-time collaborative use across interdisciplinary teams. The graphical user interface allows the user to interactively conduct the types of categorically-rooted operations discussed without any knowledge of their categorical foundations. This software is at an early stage, and currently supports only a subset of the options made possible by the StockFlow.jl framework on which it rests. But its development is progressing rapidly, and we anticipate an expansion to eventually handle a far larger set of operations.  We describe use of some of the early features of this system, bearing in mind that multiple users will commonly be using the system simultaneously.

ModelCollab provides a modal interface for adding diagram components to a Canvas used to display the diagram being assembled. Different concurrent users can be present in different modes at the same time. For example, within ``Stock" mode, the user can click on the canvas to add a Stock, and similarly for ``Flow", ``Auxiliary Variable", ``Sum Variable" modes. ``Connect" mode is used to establish links indicating dependencies, such as those for auxiliary and sum variables. The interface abstraction level sometimes exceeds that of StockFlow.jl; for example, flows within the graphical interface are shown as depending directly on other variables in the diagram, rather than only via connections with a distinct auxiliary variable. Through this interface, larger diagrams can be created; for example, Figure \ref{fig:simply_model_collab_SEIR.png} depicts an SEIR diagram built in the system.

\begin{figure}[H]
    \centering
    \includegraphics[width=0.7\columnwidth]{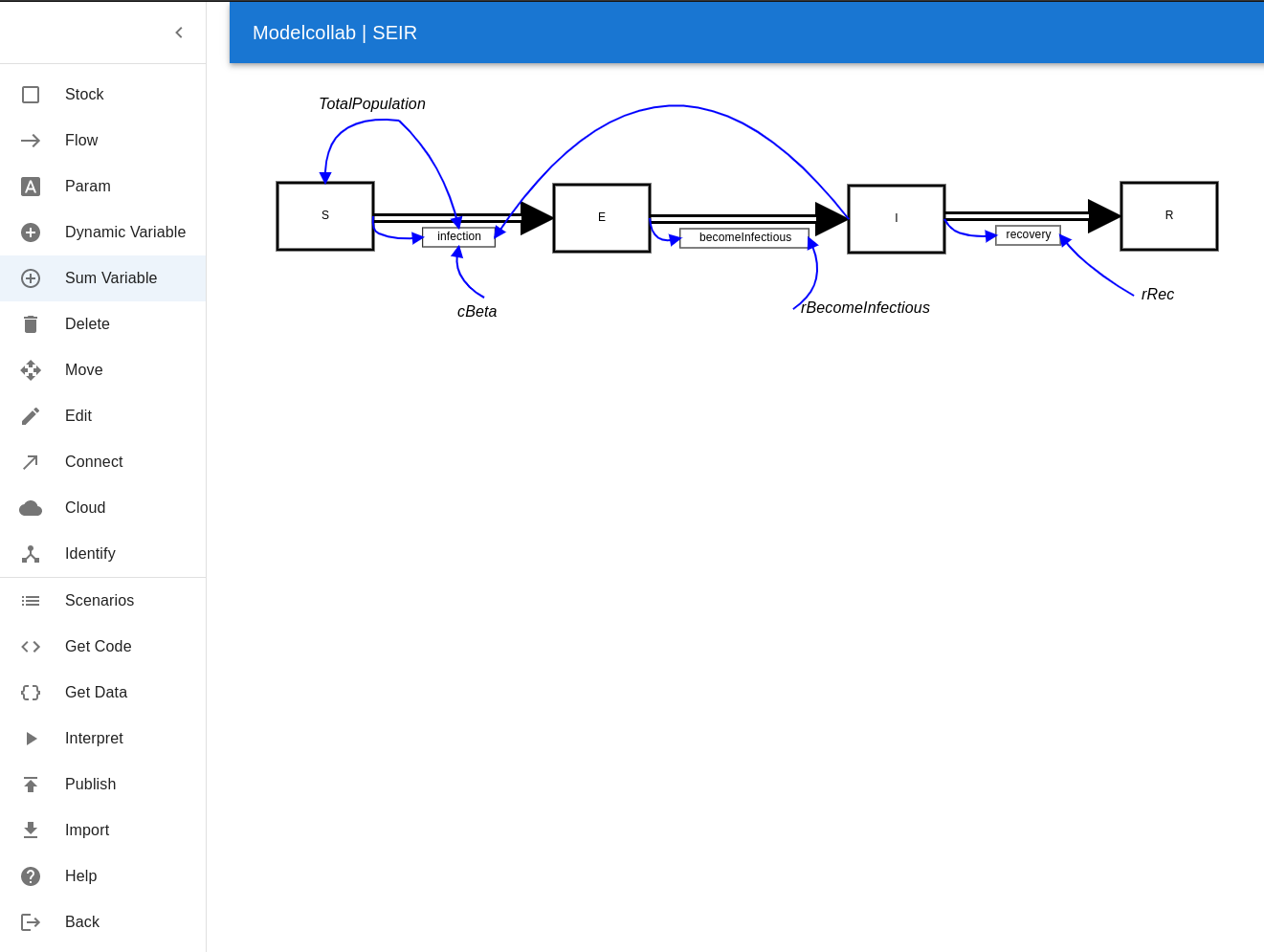}
    \caption{An SEIR diagram built in ModelCollab}
    \label{fig:simply_model_collab_SEIR.png}
\end{figure}

As in online collaborative software such as Google Docs, diagrams can be accessed on an ongoing basis by multiple parties once they have been created in ModelCollab. It is also possible to persist diagrams in other forms within the system. The interface allows export of diagrams through a menu item, with that diagram then being downloaded to the invoking user's local computer as a JSON \cite{json} file. Beyond exporting, the system offers a structured means of ``publishing" diagrams to a simple ``Diagram Library" once they have attained a sufficient level of maturity to be worth sharing. Such published diagrams can then be reused by others.

A foundationally important component of ModelCollab functionality is the ability to compose diagrams. Like other diagram assembly operations within ModelCollab, this operation is performed graphically. As a first step, the user can add previously published diagrams into the model canvas, where this imported diagram (henceforth referred to as the ``subdiagram'') is visually distinguished from the surrounding diagram being assembled, as is depicted in \ref{fig:canvas_showing_single_imported_diagram}. 

\begin{figure}[b]
    \centering
    \includegraphics[width=0.7\columnwidth]{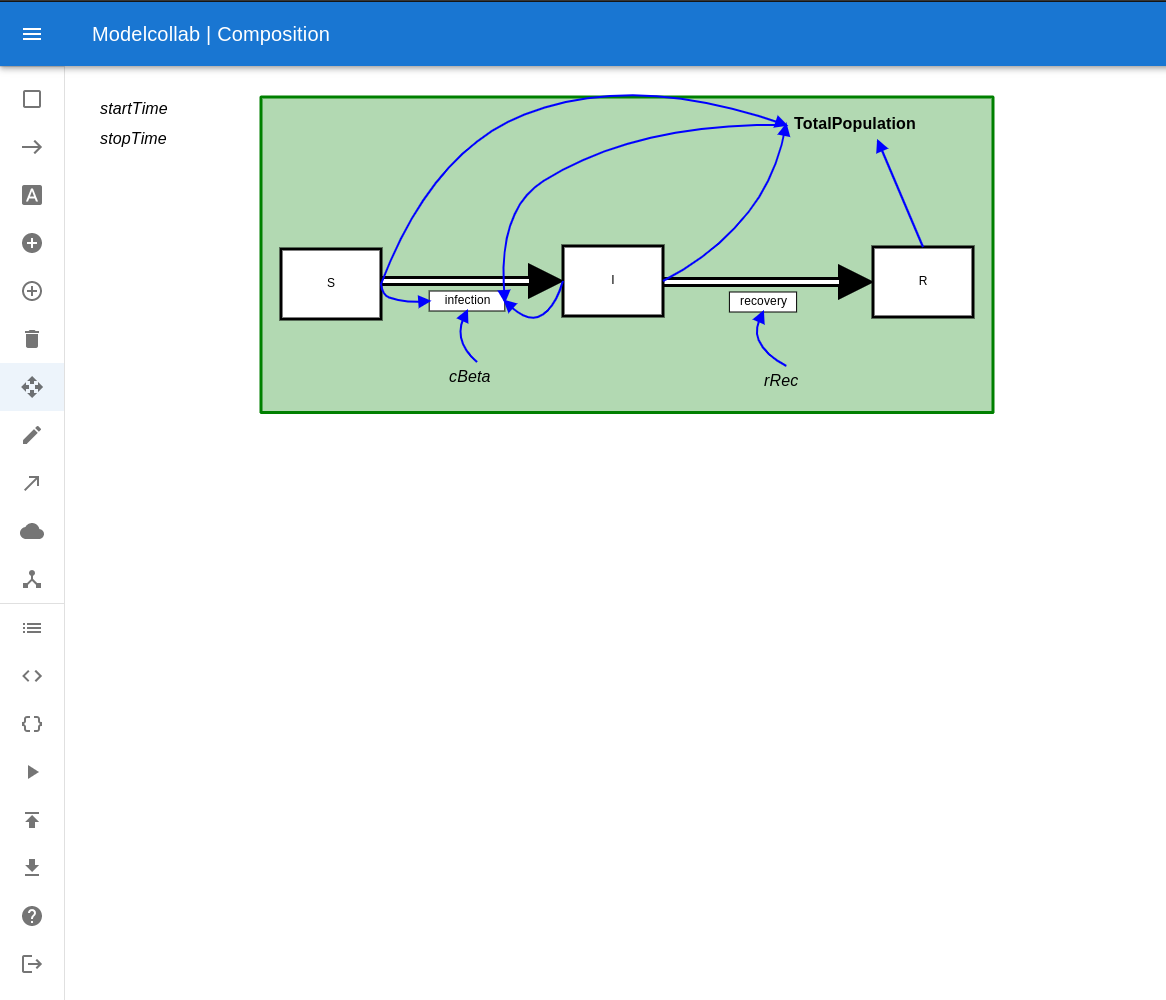}
    \caption{ModelCollab: Results of importing a single previously published diagram}
    \label{fig:canvas_showing_single_imported_diagram}
\end{figure}

More than one such published diagram can be imported, in which case each is distinguished by a different colour.  For example, Figure \ref{fig:canvas_showing_two_imported_diagrams} shows the results of importing a second subdiagram, which lies beneath the first imported subdiagram. Beyond these subdiagrams, the canvas will commonly contain a surrounding diagram.

\begin{figure}[H]
    \centering
    \includegraphics[width=0.7\columnwidth]{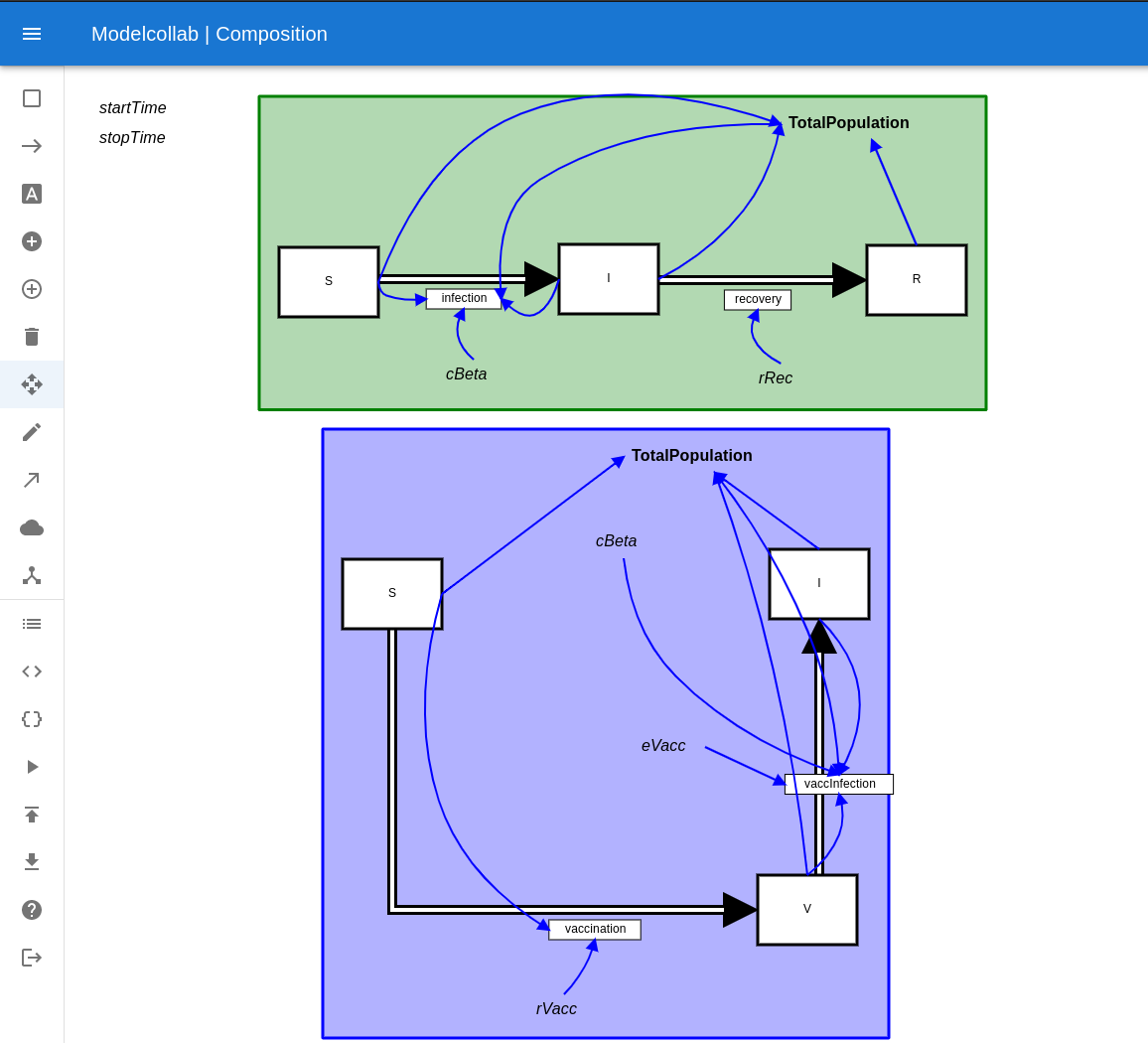}
    \caption{ModelCollab: Results of importing a second diagram}
    \label{fig:canvas_showing_two_imported_diagrams}
\end{figure}

Just after being imported, the two subdiagrams are independent of one another; while this is sometimes appropriate, often the user will recognize ways in which the process depicted in a specific imported subdiagram is coupled with the processes depicted by the surrounding diagram, or in the other imported subdiagrams. Such coupling is represented by \textit{composition} of the diagrams via the interface. Through user interface actions, a user can elect to identify (unify) a stock or sum dynamic variable with a variable of the same type in a subdiagram. Such identification of pairs of variables may be performed between variables in an outer canvas diagram and in a subdiagram, or (alternatively) between two variables in different subdiagrams.  For example, \ref{fig:single_stock_identified_between_two_imported_diagrams} graphically illustrates the results of identification of a stock V of the upper and lower diagrams.  By entering ``Identify" mode from the ModelCollab menu, the user can indicate with a pair of successive clicks the pair of stocks to be identified.

\begin{figure}[H]
    \centering
    \includegraphics[width=0.7\columnwidth]{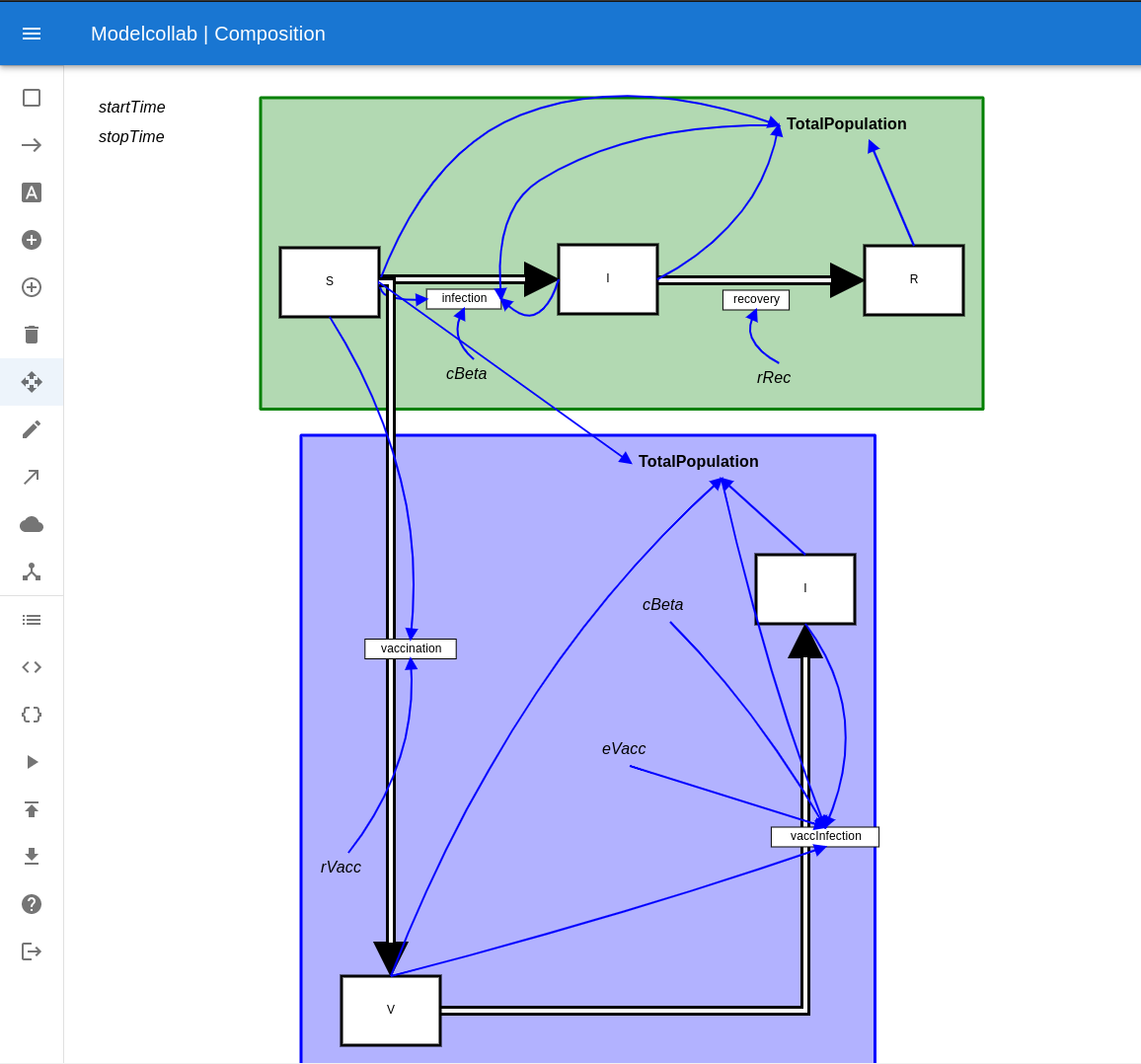}
    \caption{ModelCollab: Example of two subdiagrams composed by identifying stock S}
    \label{fig:single_stock_identified_between_two_imported_diagrams}
\end{figure}

Figure \ref{fig:S_I_N_identified_between_two_imported_diagrams} shows the results of using the ``Identify" mode of ModelCollab to identify not just additional stock I between the two diagrams, but also sum variable N.  

\begin{figure}[H]
    \centering
    \includegraphics[width=0.7\columnwidth]{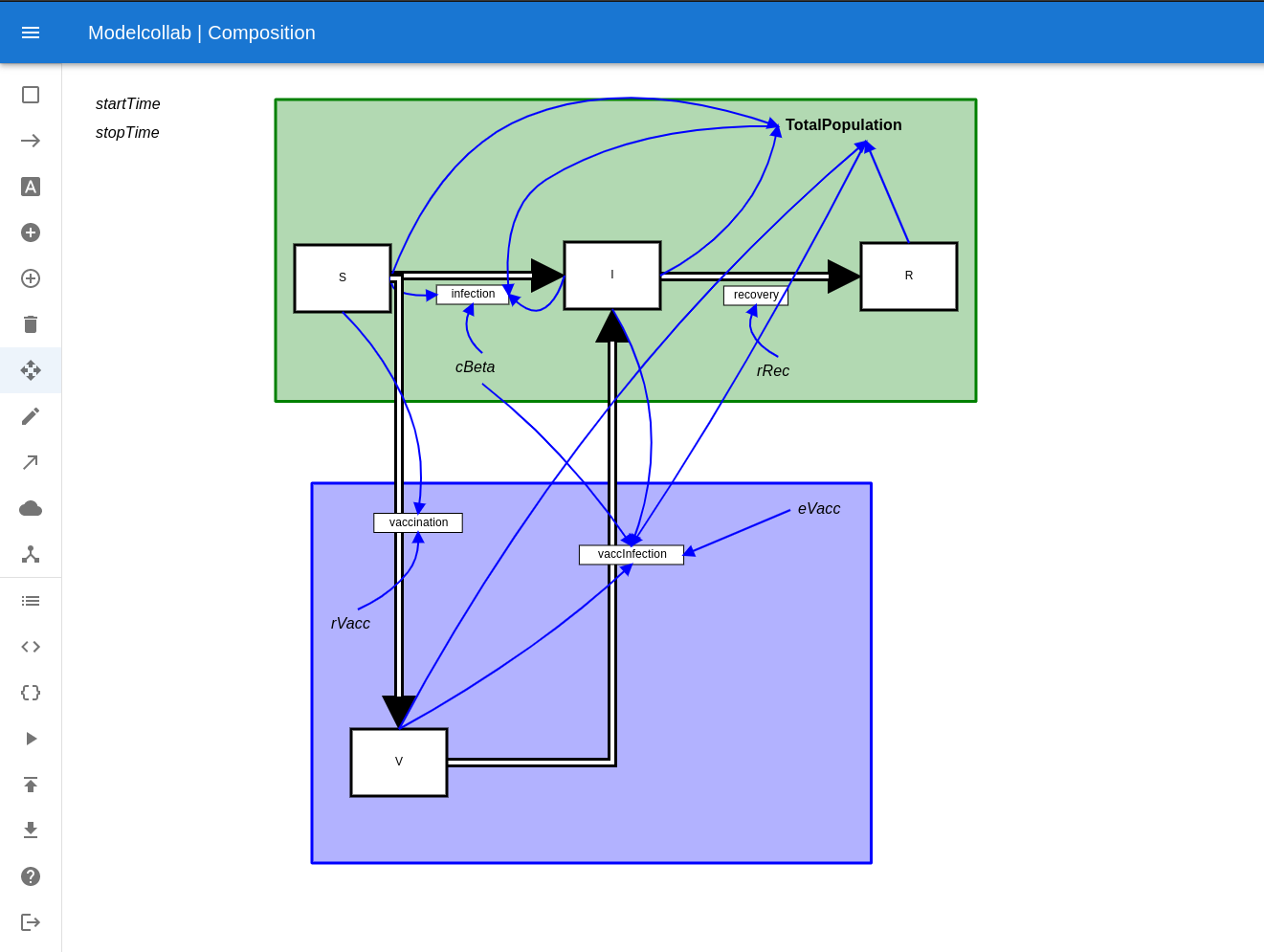}
    \caption{ModelCollab: Example of two subdiagrams composed by identifying both stocks S and I and sum variable N}
    \label{fig:S_I_N_identified_between_two_imported_diagrams}
\end{figure}

Once built, ModelCollab supports rendering the definition of a diagram in code form. Specifically, the system offers a ``Get code" menu item that produces a file containing Julia code to create the current diagram using calls to StockFlow.jl. If desired, such code could then be used to interactively manipulate the models from a Julia codebase or within a Jupyter notebook.

Beyond operating on syntactic constructs in the form of diagrams, ModelCollab provides an interface to interpret diagrams according to a menu-chosen semantics. At the time of writing, the software only supports ODE semantics (see Figure \ref{fig:SelectingODESemantics}), but implementation of the other semantics discussed in section \ref{section: semantics} is planned within the near future.

\begin{figure}[H]
    \centering
    \includegraphics[width=0.7\columnwidth]{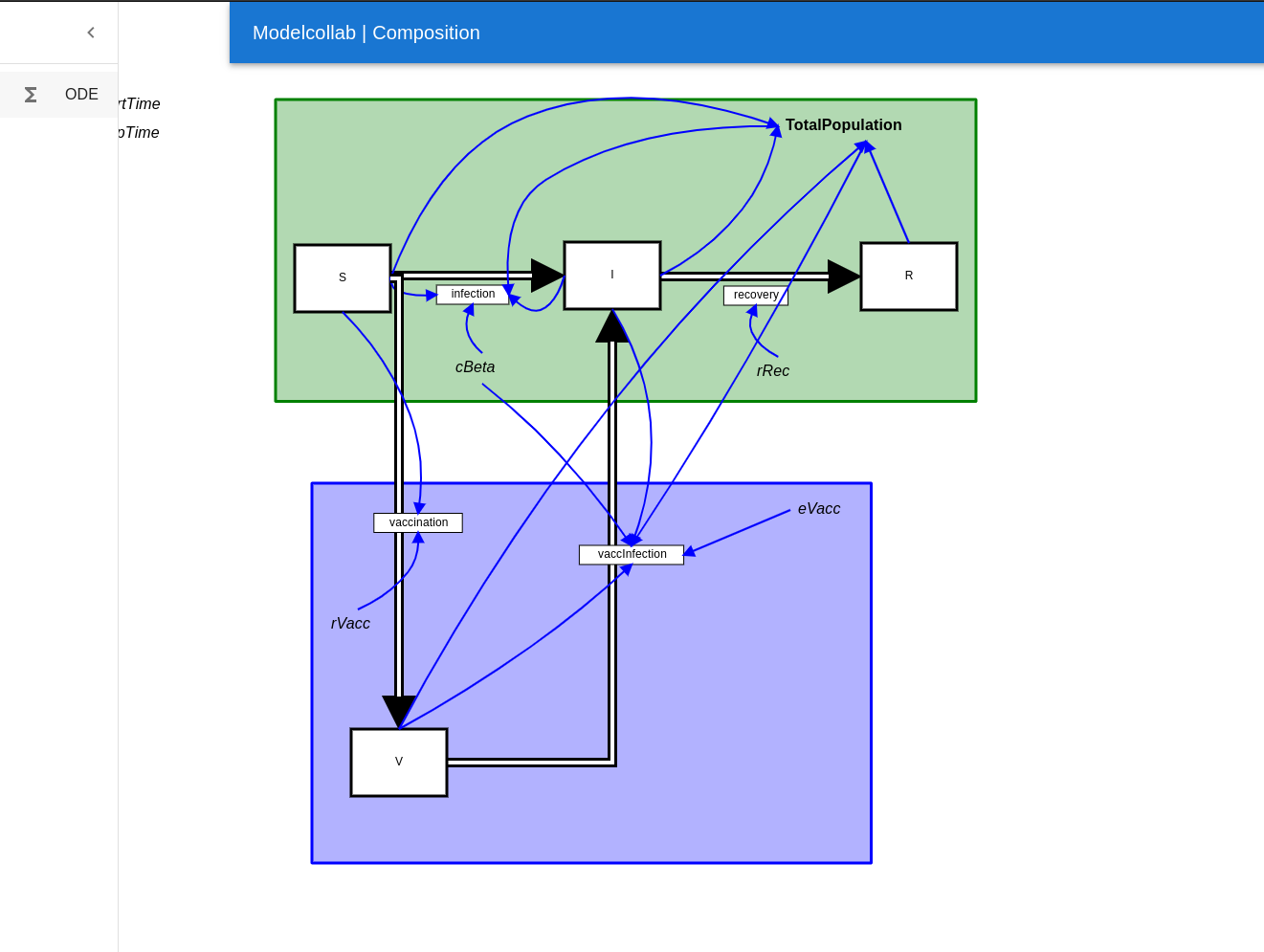}
    \caption{ModelCollab: Example of using the ``Semantics" menu to elect to interpret the diagram (the syntax) as an ODE}
    \label{fig:SelectingODESemantics}
\end{figure}

Interpretation of a diagram by ODE semantics involves numerical integration of that diagram over a user-specified timeframe. Outputs from that simulation are currently rendered and downloaded as a PNG image file; Figure \ref{fig:SimulationModelOutput} shows an example. 

\begin{figure}[H]
    \centering
    \includegraphics[width=0.7\columnwidth]{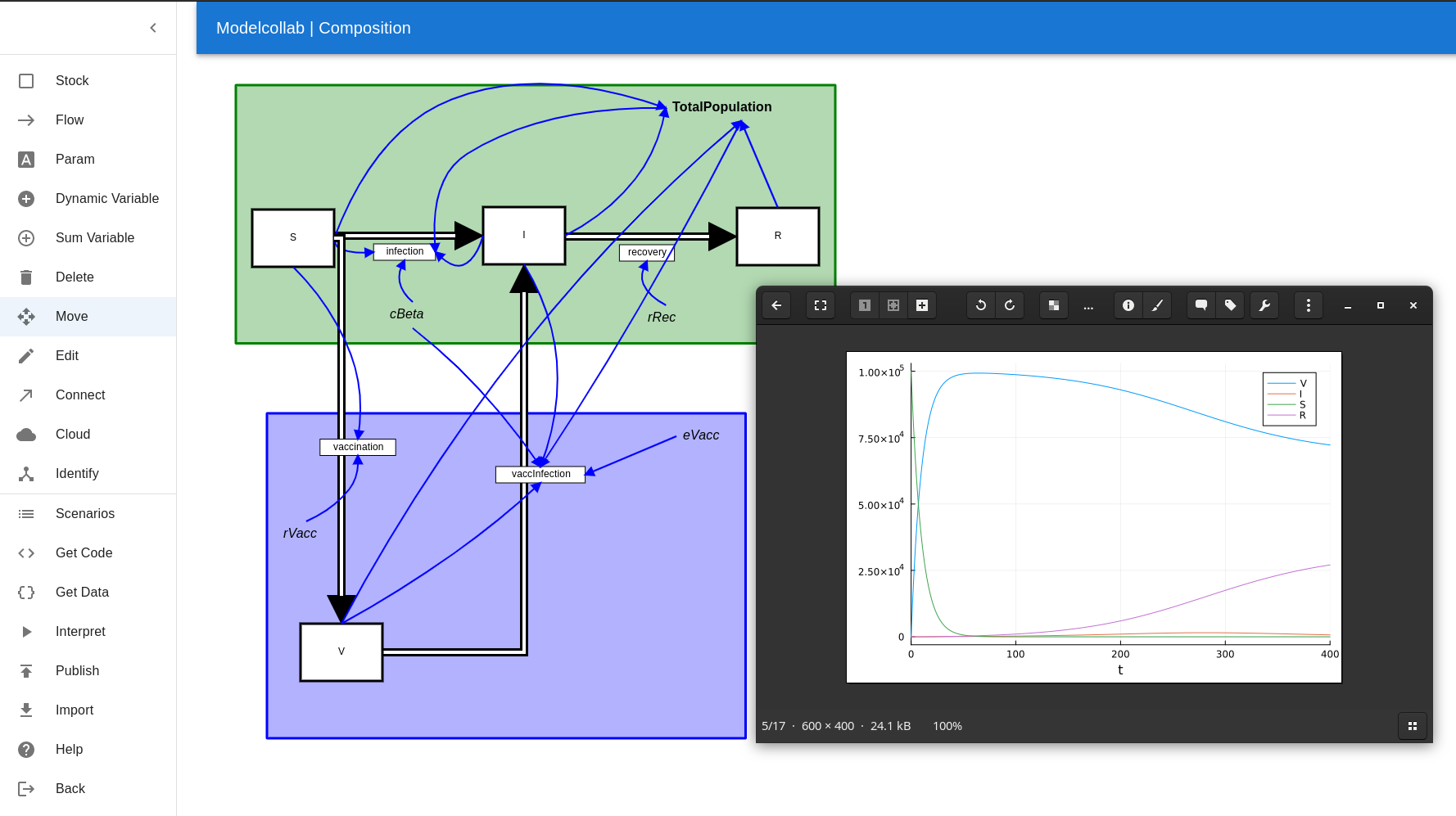}
    \caption{ModelCollab: Example of results of interpreting the composed diagrams (the syntax) with ODE semantics}
    \label{fig:SimulationModelOutput}
\end{figure}

The current ModelCollab software represents a modest step towards realizing the promise of compositional approaches for the broader modeling community. There are several priorities anticipated for rollout in the near future. Putting aside additions requiring changes for StockFlow.jl itself (some of which are discussed below), planned features include support for pullback-based stratification and additional semantic domains.  Furthermore, a scenario-based interface is being planned that will provide persistent options to interpret the model using different semantics and settings, and support other users in observing and annotating the output from semantic-based interpretation long after it is first produced.  Support is also planned for the type of seamless version control standard in many real-time collaborative systems and indications of the presence of the pointers of different concurrent users.  Finally, role-based security, authentication and authorization systems are planned to formalize permission-based enablement of diagram use, sharing and modes of interaction.

\section{Conclusion}
\label{section: conclusion}

This chapter demonstrates some of the modeling benefits secured when one takes diagrams seriously. Specifically, we have offered a brief look at some of the characteristics of categorical treatment of stock-flow diagrams, while offering a nod to other related diagrams within the diagram-centric System Dynamics modeling tradition. While our treatment has been informal and has only touched on a small subset of the consequences of a categorical foundation for diagrams, we have highlighted several benefits: the capacity to promote modularity and reuse via diagram composition, to sidestep model opacity arising from the curse of dimensionality afflicting stratified models via a modular stratification, and better supporting the needs of interdisciplinary stakeholders via the capacity to interpret the same diagrams through the lens of varying semantic domains. Some of these benefits---such as those involving semantics mapping stock-flow diagrams to system structure diagrams or causal loop diagrams---describe relationships between different types of diagrams not previously formalized.  Other outcomes of providing a firm mathematical basis for stock-flow models have been noted in passing. For example, such a formalization offers the ability to soundly transform a diagram whilst maintaining invariant its mathematical meaning, such as for optimization or parallelization of model simulation. As another example, the formalization can also allow the use of maps between diagrams that coarse-grain model structure. There are diverse other opportunities for exploiting this categorical formalization of mathematical structure.

While it offers promise, this work remains at the earliest stages of exploiting such opportunities extending from categorifying stock-flow diagrams. StockFlow.jl and ModelCollab require many important extensions to substantively address the challenges of practical modeling. Key priorities include providing support for upstream/downstream composition based on ``half-edge'' flows emerging from a diagram, supporting methods for hierarchical composition of diagrams (such as those pioneered by our colleague N.\ Meadows \cite{NicholasHierarchicalCEPHILTechReport2022}), adding full support for causal loop diagrams and system structure diagrams, and supporting the augmentation of multiple types of diagrams with dimensional information \cite{hart1995multidimensional} and use of such information in stock-flow diagram composition, stratification and additional semantic domains. There is further a need and opportunity to develop additional semantic domains for use with stock-flow models. These include those associated with different numerical simulations, such as stochastic transition systems, stochastic differential equations and difference equations, as well as those associated with computational statistics techniques. We seek to follow each such advance in the formalisms for stock-flow models by successive implementation in StockFlow.jl and ModelCollab.

At a time where health dynamic modeling is in greater demand and more needed than ever, formalizing its categorical foundation confers benefits key to addressing the team-based modeling needs of the 21st century. We enthusiastically welcome collaboration with colleagues interested in exploring opportunities to transformationally enable health modeling by tapping the power of category theory.

%Community contributions are welcomed to realize the potential of this approach.

%When the user selects a given element within the canvas, a (non-modal) properties dialog box will be shown, allowing the user to change properties of the selected item; for example, the name of an added stock can be changed. Similarly, in flow mode,

\begin{acknowledgement}
We gratefully acknowledge the extensive insights, comments and feedback from Evan Patterson of the Topos Institute.  Co-author Osgood wishes to express his appreciation of support via NSERC  via the Discovery Grants program (RGPIN 2017-04647), from the Mathematics for Public Health Network, and from SYK \& XZO.
\end{acknowledgement}

\bibliographystyle{spmpsci}
\bibliography{references}



\section*{Supplementary material}

# Appendix: Code of Examples

The code below can be reached in the GitHub repository at: [https://github.com/Xiaoyan-Li/applicationStockFlowMFPH](https://github.com/Xiaoyan-Li/applicationStockFlowMFPH).

[1]:
```julia
using StockFlow

using Catlab
using Catlab.CategoricalAlgebra
using LabelledArrays
using OrdinaryDiffEq
using Plots

using Catlab.Graphics
using Catlab.Programs
using Catlab.WiringDiagrams

using Catlab.Graphics.Graphviz: Html
using Catlab.Graphics.Graphviz
```

# 1 SEIR Model

## 1.1 Syntax: Define the SEIR Stock-Flow Diagram

Builds the stock-flow diagram of the SEIR model:

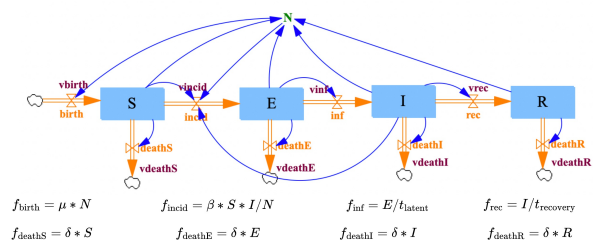

$f_{\text{birth}} = \mu * N$     $f_{\text{incid}} = \beta * S * I / N$     $f_{\text{inf}} = E / t_{\text{latent}}$     $f_{\text{rec}} = I / t_{\text{recovery}}$

$f_{\text{deathS}} = \delta * S$     $f_{\text{deathE}} = \delta * E$     $f_{\text{deathI}} = \delta * I$     $f_{\text{deathR}} = \delta * R$

[2]:
```julia
# define the functions of the auxiliary variables
f_birth(u,uN,p,t)=p.μ*uN.N(u,t)
```

```
f_incid(u,uN,p,t)= p.β*u.S*u.I/uN.N(u,t)
f_inf(u,uN,p,t)=u.E/p.tlatent
f_rec(u,uN,p,t)=u.I/p.trecovery
f_deathS(u,uN,p,t)=u.S*p.δ
f_deathE(u,uN,p,t)=u.E*p.δ
f_deathI(u,uN,p,t)=u.I*p.δ
f_deathR(u,uN,p,t)=u.R*p.δ
```

[2]: f_deathR (generic function with 1 method)

[3]:
```
seir=StockAndFlow(
    (:S=>(:birth,(:incid,:deathS),(:v_incid,:v_deathS),:N),
        :E=>(:incid,(:inf,:deathE),(:v_inf,:v_deathE),:N),
        :I=>(:inf,(:rec,:deathI),(:v_incid, :v_rec,:
 ↪v_deathI),:N),
        :R=>(:rec,:deathR,:v_deathR,:N)),
    (:birth=>:v_birth,:incid=>:v_incid,:inf=>:v_inf,:rec=>:
 ↪v_rec,:deathS=>:v_deathS,:deathE=>:v_deathE,:deathI=>:
 ↪v_deathI,:deathR=>:v_deathR),
    (:v_birth=>f_birth,:v_incid=>f_incid,:v_inf=>f_inf,:
 ↪v_rec=>f_rec,:v_deathS=>f_deathS,:v_deathE=>f_deathE,:
 ↪v_deathI=>f_deathI,:v_deathR=>f_deathR),
    (:N=>(:v_birth,:v_incid))
);
```

We can also plot the SEIR stock-flow diagram using the function Graph() as below. This stock-flow diagram is the same as the upper figure. Although the software plotted diagrams may be not as good as the upper figure, it will be improved in the future.

[4]: `Graph(seir)`

[4]:

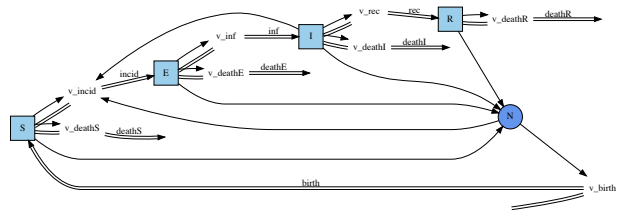

## 1.2 Semantics: ODEs

### 1.2.1 Define the constant parameter values and the initial values for the stocks

Time unit: Month

```
[5]: # define parameter values and initial values of stocks
     # define constant parameters
     p_measles = LVector(
         β=49.598, μ=0.03/12, δ=0.03/12, tlatent=8.0/30,
         ↪trecovery=5.0/30
     )
     # define initial values for stocks
     u0_measles = LVector(
         S=90000.0-930.0, E=0.0, I=930.0, R=773545.0
     );
```

### 1.2.2 Solve the ODEs and plot the results

```
[6]: prob_measles = ODEProblem(vectorfield(seir),u0_measles,(0.
     ↪0,120.0),p_measles);
     sol_measles = solve(prob_measles,Tsit5(),abstol=1e-8);
     plot(sol_measles)
```

[6]:

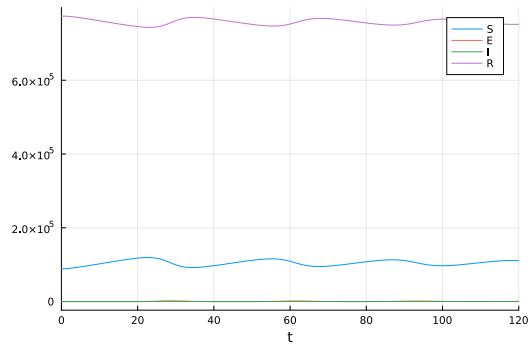

## 1.3 Semantics: Causal Loop Diagrams

**Convert the SEIR stock-flow diagram to a causal loop diagram:**

```
[7]: seir_causalLoop=convertToCausalLoop(seir);
```

```
[8]: Graph(seir_causalLoop)
```

[8]:

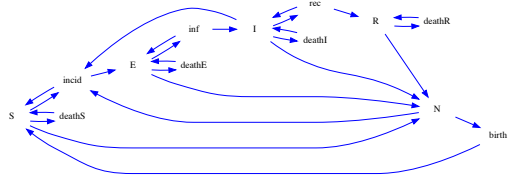

## 1.4 Semantics: System Structure Diagrams

**Convert the SEIR stock-flow diagram to a system structure diagram:**

[9]: 
```
seir_structure=convertStockFlowToSystemStructure(seir);
```

[10]: 
```
Graph(seir_structure)
```

[10]:

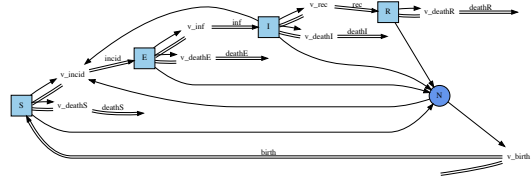

# 2 Composition

## 2.1 Open population SVE model

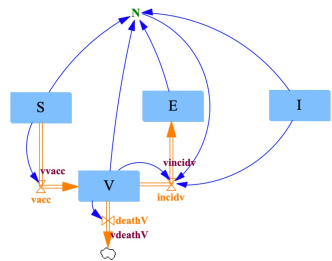

$f_{\text{vacc}} = \alpha * S \qquad f_{\text{deathV}} = \delta * V$

$f_{\text{incidv}} = \beta * (1 - e) * V * I/N$

```
[11]: f_vacc(u,uN,p,t)=u.S*p.α
      f_deathV(u,uN,p,t)=u.V*p.δ
      f_incidv(u,uN,p,t)=p.β*u.V*u.I*(1.0-p.e)/uN.N(u,t)

      sve=StockAndFlow(
          (:S=>(:F_NONE,:vacc,:v_vacc,:N),
              :V=>(:vacc,(:deathV,:incidv),(:v_deathV, :v_incidv),:
       ↪N),
              :E=>(:incidv,:F_NONE,:V_NONE,:N),
              :I=>(:F_NONE,:F_NONE,:v_incidv,:N)),
          (:vacc=>:v_vacc,:deathV=>:v_deathV, :incidv=>:v_incidv),
          (:v_vacc=>f_vacc,:v_deathV=>f_deathV,:
       ↪v_incidv=>f_incidv),
          (:N=>:v_incidv)
      )

      Graph(sve)
```

[11]: 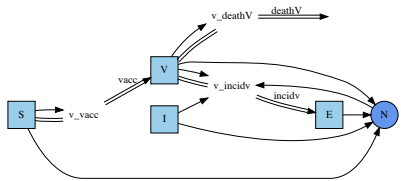

## 2.2 Composition

### 2.2.1 Define the composition UWD-algebra

```
[12]: seirv_uwd = @relation (S,E,I) begin
          seir(S,E,I)
          sve(S,E,I)
      end;
      display_uwd(seirv_uwd)
```

[12]: 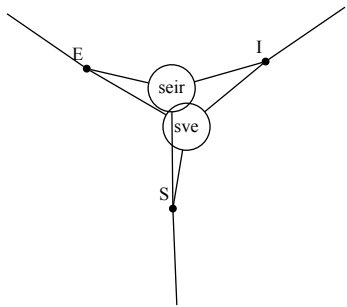

5

### 2.2.2 Define the common parts (feet) to be composed of

[13]:
```
footS=foot(:S, :N, :S=>:N)
Graph(footS;schema="C0")
```

[13]:

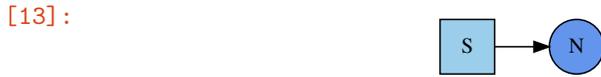

[14]:
```
footE=foot(:E, :N, :E=>:N)
Graph(footE;schema="C0")
```

[14]:

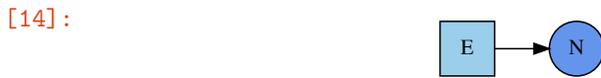

[15]:
```
footI=foot(:I, :N, :I=>:N)
Graph(footI;schema="C0")
```

[15]:

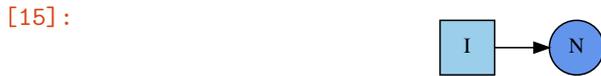

### 2.2.3 Composition

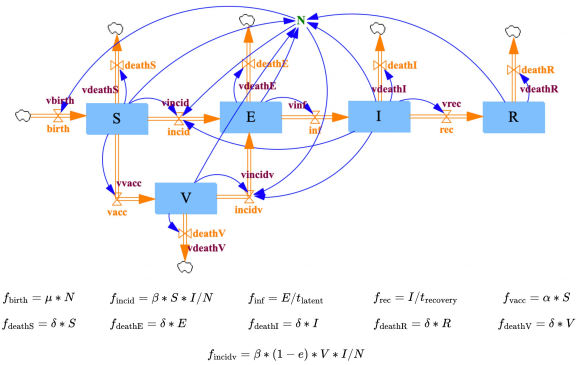

$f_{\text{birth}} = \mu * N$  $\quad f_{\text{incid}} = \beta * S * I/N$  $\quad f_{\text{inf}} = E/t_{\text{latent}}$  $\quad f_{\text{rec}} = I/t_{\text{recovery}}$  $\quad f_{\text{vacc}} = \alpha * S$

$f_{\text{deathS}} = \delta * S$  $\quad f_{\text{deathE}} = \delta * E$  $\quad f_{\text{deathI}} = \delta * I$  $\quad f_{\text{deathR}} = \delta * R$  $\quad f_{\text{deathV}} = \delta * V$

$f_{\text{incidv}} = \beta * (1 - e) * V * I/N$

```
[16]: # open SEIR and SVE stock & flow diagram with the feet
      ↪defined before
      open_seir=Open(seir, footS, footE, footI)
      open_sve=Open(sve, footS, footE, footI)
      # Compose those two models according the UWD-algebra
      open_seirv = oapply(seirv_uwd, [open_seir, open_sve])
      # the composed closed stock & flow diagram is the apex of
      ↪the open stock & flow diagram
      seirv = apex(open_seirv)
      Graph(seirv)
```

[16]:

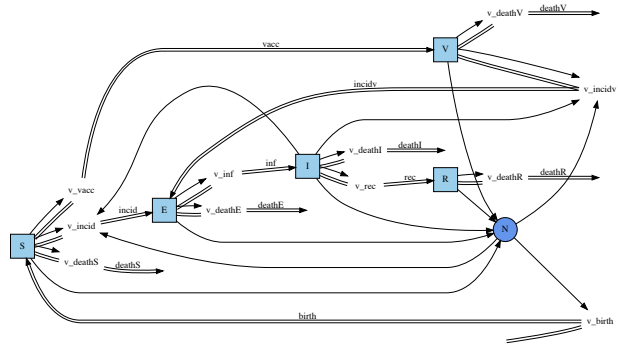

### 2.2.4 Solve the ODEs of the SEIRV model

Time Unit: days

```
[17]: p_seirv = LVector(
          β=49.598/30, μ=0.03/365, δ=0.03/365, tlatent=8.0,
      ↪trecovery=5.0, α=0.01, e=0.9
      )
      # define initial values for stocks
      u0_seirv = LVector(
          S=10000.0-1.0, E=0.0, I=1.0, R=0.0, V=0.0
      );

      prob_seirv = ODEProblem(vectorfield(seirv),u0_seirv,(0.0,100.
      ↪0),p_seirv);
      sol_seirv = solve(prob_seirv,Tsit5(),abstol=1e-8);
      plot(sol_seirv)
```

[17]:

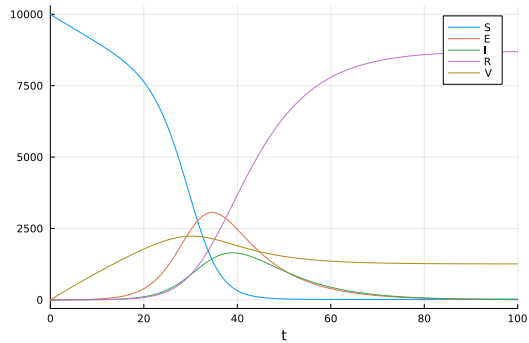

# 3 Introduction of Stratification

## 3.1 Stratification

Stratification is based on system structure diagrams. Stratified stock-flow diagrams are then generated by assigning a function to each auxiliary variable of the stratified system structure diagram.

We generate a stratified system structure diagram $S_\text{stratified}$ by taking the pullback of an aggregate population infectious disease system structure diagram ($S_\text{aggregate}$ and a strata system structure diagram $S_\text{strata}$, which are both mapped to the same system structure diagram $S_\text{type}$:

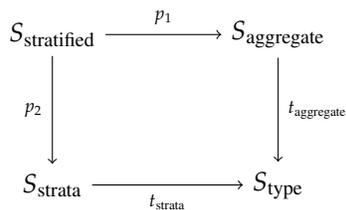

The map $t_\text{aggregate} \colon S_\text{aggregate} \to S_\text{type}$ assigns a type to each part of the system structure diagram $S_\text{aggregate}$. The allowed types are described by the type system $S_\text{type}$. Similarly, the map $t_\text{strata}$ assigns a type to each part of the system structure diagram $S_\text{strata}$.

In applications, we can define different system structure diagrams $S_\text{aggregate}$ to represent different infectious disease models (or more generally, compartmental models), e.g., the SEIR (Susceptible-Exposed-Infectious-Recovered) model, SIR (Susceptible-Infectious-Recovered) model, SIS (Susceptible-Infectious-Susceptible) model, etc. Similarly, the strata system structure diagrams $S_\text{strata}$ are

8

consist of multiple different strata models, e.g., the age strata, the sex strata, the region strata, etc. Then, a stratified model can be build by calculating the pullback of a disease model $S_{\text{aggregate}}$ and a strata model $S_{\text{strata}}$. For example, we can generate a SEIR age stratified system structure diagram by calculating the pullback of the SEIR disease model and the age strata model.

In this project, we built two strata models (age strata model and sex strata model), and two disease models (SEIR model and SIS model) as examples. Four stratified models were generated by calculating the pullback of the four different combination of the two strata models and two disease models. Those four models are an age stratified SEIR model, an age stratified SIS model, a sex stratified SEIR model and a sex stratified SIS model. These are shown in Figure 14.

### 3.1.1 Define the typed system structure diagram

First, we need to define the type system $S_{\text{type}}$. In this system structure diagram, each component represents a type.

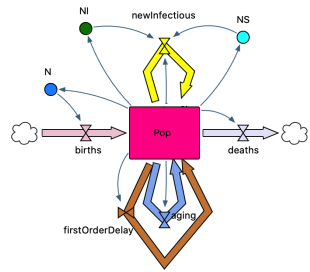

The type system used in this project is shown above. From this we can see there is only one type of stock, which represents the population. There are five flows from this stock to itself, which represent five different flow types in the infectious disease models under consideration:

- the birth flow
- the death flow
- the incident infections flow
- the aging flow which represents the flow of aging from one age group to the older group
- the first order delay flow

There should also be five auxiliary variables corresponding to these five flows. For simplicity our system structure diagram does not show these auxiliary variables.

The system structure diagram $S_{\text{type}}$ shown above also has three sum auxiliary variables:

- N: the total population of the whole model
- NS: the total population of a specific subgroup
- NI: the total count of infectious people across a specific subgroup

Finally, each link in $S_{\text{type}}$ represents a type of link. However, since we can easily determine a link's type by the types of its source and target, we did not label the links' types by colours.

We define the typed stock-flow diagram:

```
[18]: S_type=StockAndFlowStructure(
          (:Pop=>((:births, :newInfectious, :firstOrderDelay, :
      ↪aging),(:deaths, :newInfectious, :firstOrderDelay, :
      ↪aging),(:v_deaths, :v_newInfectious, :v_firstOrderDelay, :
      ↪v_aging),(:NS,:NI,:N))),
          (:deaths=>:v_deaths, :births=>:v_births, :
      ↪newInfectious=>:v_newInfectious, :firstOrderDelay=>:
      ↪v_firstOrderDelay, :aging=>:v_aging),
          (:N=>:v_births, :NI=>:v_newInfectious, :NS=>:
      ↪v_newInfectious)
      );
```

We can also plot the typed diagrams. The typed system structure diagram is the same as the upper figure. At first, we need to define the function of plotting out the typed diagrams and some needed help functions:

```
[19]: # Functions for graphing typed stock & flow diagrams
      colors_vflow = ["antiquewhite4","antiquewhite", "gold",
      ↪"saddlebrown", "slateblue", "blueviolet", "olive"]
      colors_s = ["deeppink","darkorchid","darkred","coral"] # red
      ↪series
      colors_sv = ["cornflowerblue","cyan4","cyan","chartreuse"] #
      ↪green and blue series

      flatten(fname::Symbol) = "$fname"

      function flatten(fname::Tuple)
          names = split(replace(string(fname), "("=>"", ")"=>"", ":
      ↪"=>""), ",")
          for i in 1:length(names)
              name = strip(names[i])
              if name[1:2] == "id"
```

```
            continue
        end
        return name
    end
    return "id"
end

def_stock(typed_StockFlow::ACSetTransformation, colors) =
  (p,s) -> ("s$s", Attributes(:label=>sname(p,s) isa Tuple␣
 ↪where T ? Html(replace(string(sname(p,s)), ":"=>"", "," =>␣
 ↪"<BR/>", "("=>"", ")"=>"")) : "$(sname(p,s))",
                                              :shape=>"square",
                                              :color=>"black",
                                              :style=>"filled",
                                                                :
 ↪fillcolor=>colors[typed_StockFlow[:S](s)]))

def_auxiliaryV(typed_StockFlow::ACSetTransformation, colors)=
  (p, v) -> ("v$v", Attributes(:label=>vname(p,v) isa Tuple␣
 ↪where T ? Html(replace(string(vname(p,v)), ":"=>"", "," =>␣
 ↪"<BR/>", "("=>"", ")"=>"")) : "$(vname(p,v))",
                                                                :
 ↪shape=>"plaintext",
                                                                :
 ↪fontcolor=>colors[typed_StockFlow[:F](incident(p,v,:fv)...
 ↪)]))


def_auxiliaryV(colors = colors_vflow)=
  (p, v) -> ("v$v", Attributes(:label=>vname(p,v) isa Tuple␣
 ↪where T ? Html(replace(string(vname(p,v)), ":"=>"", "," =>␣
 ↪"<BR/>", "("=>"", ")"=>"")) : "$(vname(p,v))",
                                                                :
 ↪shape=>"plaintext",
                                                                :
 ↪fontcolor=>colors[incident(p,v,:fv)...]))



def_sumV(typed_StockFlow::ACSetTransformation, colors) =
  (p, sv) -> ("sv$sv", Attributes(:label=>svname(p,sv) isa␣
 ↪Tuple where T ? Html(replace(string(svname(p,sv)), ":
 ↪"=>"", "," => "<BR/>", "("=>"", ")"=>"")) :␣
 ↪"$(svname(p,sv))",
```

```
                                          :shape=>"circle",
                                          :color=>"black",
                                          :
 ↪fillcolor=>colors[typed_StockFlow[:SV](sv)],
                                          :style=>"filled"))


def_flow_V(typed_StockFlow::ACSetTransformation, colors)=
  (p, us, ds, v, f) -> begin
    labelfontsize = "6"
    colorType = colors[typed_StockFlow[:F](f)]
    color = "$colorType"*":invis:"*"$colorType"
    arrowhead = "none"
    splines = "ortho"
    return ([us, "v$v"],Attributes(:label=>"", :
 ↪labelfontsize=>labelfontsize, :color=>color, :
 ↪arrowhead=>arrowhead, :splines=>splines)),
            (["v$v", ds],Attributes(:
 ↪label=>Html(flatten(fname(p,f))), :
 ↪labelfontsize=>labelfontsize, :color=>color, :
 ↪splines=>splines))
end

def_flow_noneV(typed_StockFlow::ACSetTransformation, colors)=
  (p, us, ds, f) -> begin
     colorType = colors[typed_StockFlow[:F](f)]
     color = "$colorType"*":invis:"*"$colorType"
     ([us, ds],Attributes(:label=>Html(flatten(fname(p,f))),␣
 ↪:labelfontsize=>"6", :color=>color))
end

def_flow_V(colors = colors_vflow)=
  (p, us, ds, v, f) -> begin
    labelfontsize = "6"
    colorType = colors[f]
    color = "$colorType"*":invis:"*"$colorType"
    arrowhead = "none"
    splines = "ortho"
    return ([us, "v$v"],Attributes(:label=>"", :
 ↪labelfontsize=>labelfontsize, :color=>color, :
 ↪arrowhead=>arrowhead, :splines=>splines)),
```

```
           (["v$v", ds],Attributes(:
 label=>Html(flatten(fname(p,f))), :
 labelfontsize=>labelfontsize, :color=>color, :
 splines=>splines))

end

def_flow_noneV(colors = colors_vflow)=
  (p, us, ds, f) -> begin
     colorType = colors[f]
     color = "$colorType"*":invis:"*"$colorType"
     ([us, ds],Attributes(:label=>Html(flatten(fname(p,f))),
 :labelfontsize=>"6", :color=>color))
end

# Plot the typed stock & flow diagram
Graph_typed(typed_StockFlow::ACSetTransformation,
 colors_vflow = colors_vflow, colors_s = colors_s,
 colors_sv = colors_sv; schema::String="C", type::
 String="SFVL", rd::String="LR") =
 Graph(dom(typed_StockFlow),
    make_stock = def_stock(typed_StockFlow, colors_s),
 make_auxiliaryV=def_auxiliaryV(typed_StockFlow,
 colors_vflow), make_sumV=def_sumV(typed_StockFlow,
 colors_sv),
    make_flow_V=def_flow_V(typed_StockFlow, colors_vflow),
 make_flow_noneV=def_flow_noneV(typed_StockFlow,
 colors_vflow),schema=schema, type=type, rd=rd
);
```

Now, let's plot the typed diagram:

[20]: ```
Graph_typed(id(S_type))
```

[20]:

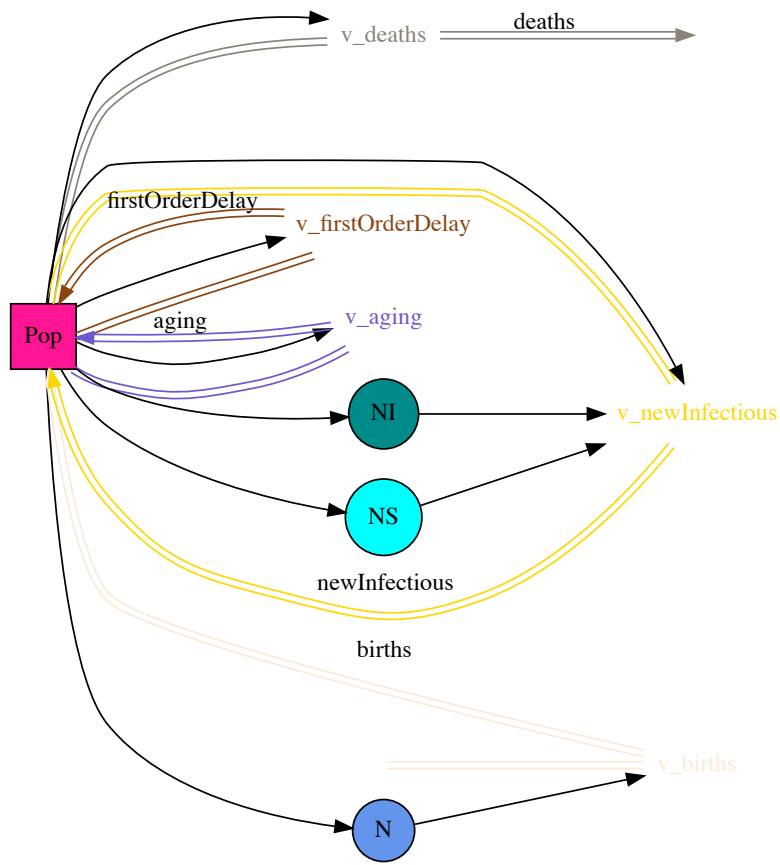

```
[21]:  # eliminate the attribute of name to enable pass the natural
       ↪check
       S_type = map(S_type, Name=name->nothing);
```

Since the typed disease model (e.g., SEIR model) and typed strata model (e.g., age strata, sex strata) are homomorphisms (with ACSetTransformation data structure) from disease (e.g., SEIR) system structure diagrams or the strata system structure diagrams to the type system structure diagram. To define the ACSetTransformation easily, we extract the parts of each components of the type system structure diagram as follows:

```
[22]:  s, = parts(S_type, :S)
       sv_N, sv_NI, sv_NS = parts(S_type, :SV)
       lsn_NS, lsn_NI, lsn_N = parts(S_type, :LS)
```

```
f_deaths, f_births, f_newInfectious, f_firstOrderDelay,
    ↪f_aging = parts(S_type, :F)
i_births, i_newInfectious, i_firstOrderDelay, i_aging =
    ↪parts(S_type, :I)
o_deaths, o_newInfectious, o_firstOrderDelay, o_aging =
    ↪parts(S_type, :O)
v_deaths, v_births, v_newInfectious, v_firstOrderDelay,
    ↪v_aging = parts(S_type, :V)
lv_deaths, lv_newInfectious, lv_firstOrderDelay, lv_aging =
    ↪parts(S_type, :LV)
lsv_N_births, lsv_NI_newInfectious, lsv_NS_newInfectious =
    ↪parts(S_type, :LSV);
```

### 3.1.2 Define the typed infectious disease aggregate population system structure diagrams

**Define the typed SEIR aggregate population system structure diagram**

[23]:
```
# the id flow of each stock maps to the aging flow in the
    ↪S_type daigram in this example
S_seir=StockAndFlowStructure(
    (:S=>((:birth,:id_S),(:incid,:deathS,:id_S),(:v_incid,:
    ↪v_deathS,:v_idS),(:N,:NS)),
        :E=>((:incid,:id_E),(:inf,:deathE,:id_E),(:v_inf,:
    ↪v_deathE,:v_idE),(:N,:NS)),
        :I=>((:inf,:id_I),(:rec,:deathI,:id_I),(:v_incid, :
    ↪v_rec,:v_deathI,:v_idI),(:N,:NS,:NI)),
        :R=>((:rec,:id_R),(:deathR,:id_R),(:v_deathR,:
    ↪v_idR),(:N,:NS))),
    (:birth=>:v_birth,:incid=>:v_incid,:inf=>:v_inf,:rec=>:
    ↪v_rec,:deathS=>:v_deathS,:deathE=>:v_deathE,:deathI=>:
    ↪v_deathI,:deathR=>:v_deathR,
        :id_S=>:v_idS,:id_E=>:v_idE,:id_I=>:v_idI,:id_R=>:
    ↪v_idR),
    (:N=>:v_birth,:NS=>:v_incid,:NI=>:v_incid)
);
```

define the typed SEIR model, which is a ACSetTransformation map from the $S_{\text{seir}}$ model to the $S_{\text{type}}$ model.

15

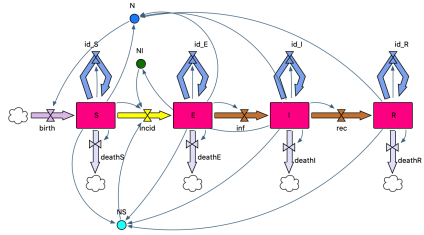

[24]:
```
t_seir=ACSetTransformation(S_seir, S_type,
  S = [s, s, s, s],
  SV = [sv_N, sv_NS, sv_NI],
  LS = [lsn_N, lsn_NS, lsn_N, lsn_NS, lsn_N, lsn_NS, lsn_NI,
  lsn_N, lsn_NS],
  F = [f_births, f_newInfectious, f_firstOrderDelay,
  f_firstOrderDelay, f_deaths, f_deaths, f_deaths, f_deaths,
  f_aging, f_aging, f_aging, f_aging],
  I = [i_births, i_aging, i_newInfectious, i_aging,
  i_firstOrderDelay, i_aging, i_firstOrderDelay, i_aging],
  O = [o_newInfectious, o_deaths, o_aging,
  o_firstOrderDelay, o_deaths, o_aging, o_firstOrderDelay,
  o_deaths, o_aging, o_deaths, o_aging],
  V = [v_births, v_newInfectious, v_firstOrderDelay,
  v_firstOrderDelay, v_deaths, v_deaths, v_deaths, v_deaths,
  v_aging, v_aging, v_aging, v_aging],
  LV = [lv_newInfectious, lv_deaths, lv_aging,
  lv_firstOrderDelay, lv_deaths, lv_aging, lv_newInfectious,
  lv_firstOrderDelay, lv_deaths, lv_aging, lv_deaths,
  lv_aging],
  LSV = [lsv_N_births, lsv_NS_newInfectious,
  lsv_NI_newInfectious],
  Name = name -> nothing
)

@assert is_natural(t_seir)
Graph_typed(t_seir)
```

[24]:

16

### Define the typed SIS aggregate population system structure diagram

```
[25]: S_sis=StockAndFlowStructure(
      (:S=>((:births,:id_S,:newRecovery),(:deathS,:id_S,:
      ↪newInfectious),(:v_deathS,:v_idS,:v_newInfectious),(:N,:
      ↪NS)),
            :I=>((:newInfectious,:id_I),(:deathI,:id_I,:
      ↪newRecovery),(:v_deathI,:v_idI,:v_newRecovery),(:N,:NS,:
      ↪NI))),
         (:births=>:v_births,:deathS=>:v_deathS,:deathI=>:
      ↪v_deathI,:newInfectious=>:v_newInfectious,:newRecovery=>:
      ↪v_newRecovery,:id_S=>:v_idS,:id_I=>:v_idI),
         (:N=>:v_births, :NI=>:v_newInfectious, :NS=>:
      ↪v_newInfectious)
         )

t_sis=ACSetTransformation(S_sis, S_type,
  S = [s, s],
  SV = [sv_N, sv_NI, sv_NS],
  LS = [lsn_N, lsn_NS, lsn_N, lsn_NS, lsn_NI],
```

```
  F = [f_births, f_deaths, f_deaths, f_newInfectious,
 ↪f_firstOrderDelay, f_aging, f_aging],
  I = [i_births, i_aging, i_firstOrderDelay,
 ↪i_newInfectious, i_aging],
  O = [o_deaths, o_aging, o_newInfectious, o_deaths,
 ↪o_aging, o_firstOrderDelay],
  V = [v_births, v_deaths, v_deaths, v_newInfectious,
 ↪v_firstOrderDelay, v_aging, v_aging],
  LV = [lv_deaths, lv_aging, lv_newInfectious, lv_deaths,
 ↪lv_aging, lv_firstOrderDelay],
  LSV = [lsv_N_births, lsv_NI_newInfectious,
 ↪lsv_NS_newInfectious],
  Name = name -> nothing
)

@assert is_natural(t_sis)
Graph_typed(t_sis)
```

[25]:

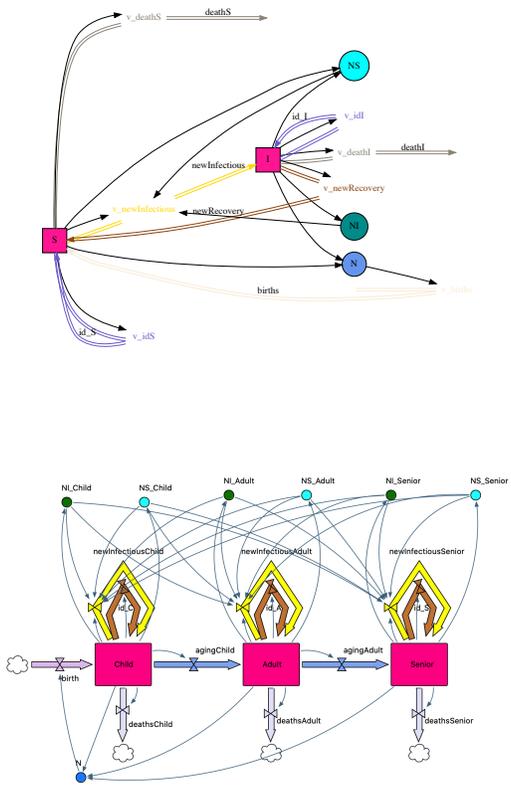

### 3.1.3 Define the typed strata system structure diagrams

**Define the aging strata system structure diagram by three age groups: Child, Adult and Senior**

[26]:
```
S_age_strata=StockAndFlowStructure(
    (:Child=>((:births,:newInfectiousChild,:id_C),(:
 ↪deathsChild,:newInfectiousChild,:agingChild,:id_C),(:
 ↪v_deathsChild,:v_newInfectiousChild,:v_agingChild,:
 ↪v_id_C),(:NI_Child,:NS_Child,:N)),
     :Adult=>((:agingChild,:newInfectiousAdult,:id_A),(:
 ↪deathsAdult,:newInfectiousAdult,:agingAdult,:id_A),(:
 ↪v_deathsAdult,:v_newInfectiousAdult,:v_agingAdult,:
 ↪v_id_A),(:NI_Adult,:NS_Adult,:N)),
     :Senior=>((:agingAdult,:newInfectiousSenior,:id_S),(:
 ↪deathsSenior,:newInfectiousSenior,:id_S),(:v_deathsSenior,:
 ↪v_newInfectiousSenior,:v_id_S),(:NI_Senior,:NS_Senior,:
 ↪N))),
    (:births=>:v_births,:newInfectiousChild=>:
 ↪v_newInfectiousChild,:newInfectiousAdult=>:
 ↪v_newInfectiousAdult,:newInfectiousSenior=>:
 ↪v_newInfectiousSenior,
     :id_C=>:v_id_C,:id_A=>:v_id_A,:id_S=>:v_id_S,
     :agingChild=>:v_agingChild,:agingAdult=>:
 ↪v_agingAdult,
     :deathsChild=>:v_deathsChild,:deathsAdult=>:
 ↪v_deathsAdult,:deathsSenior=>:v_deathsSenior),
    (:N=>:v_births, :NI_Child=>(:v_newInfectiousChild,:
 ↪v_newInfectiousAdult,:v_newInfectiousSenior), :NS_Child=>(:
 ↪v_newInfectiousChild,:v_newInfectiousAdult,:
 ↪v_newInfectiousSenior),
                :NI_Adult=>(:v_newInfectiousChild,:
 ↪v_newInfectiousAdult,:v_newInfectiousSenior), :NS_Adult=>(:
 ↪v_newInfectiousChild,:v_newInfectiousAdult,:
 ↪v_newInfectiousSenior),
                :NI_Senior=>(:v_newInfectiousChild,:
 ↪v_newInfectiousAdult,:v_newInfectiousSenior), :
 ↪NS_Senior=>(:v_newInfectiousChild,:v_newInfectiousAdult,:
 ↪v_newInfectiousSenior))
)

t_age_strata=ACSetTransformation(S_age_strata, S_type,
  S = [s,s,s],
  SV = [sv_N, sv_NI, sv_NS, sv_NI, sv_NS, sv_NI, sv_NS],
```

```julia
    LS = [lsn_NI, lsn_NS, lsn_N, lsn_NI, lsn_NS, lsn_N,
    ↪lsn_NI, lsn_NS, lsn_N],
    F = [f_births, f_newInfectious, f_newInfectious,
    ↪f_newInfectious, f_firstOrderDelay, f_firstOrderDelay,
    ↪f_firstOrderDelay, f_aging, f_aging, f_deaths, f_deaths,
    ↪f_deaths],
    I = [i_births, i_newInfectious, i_firstOrderDelay,
    ↪i_aging, i_newInfectious, i_firstOrderDelay, i_aging,
    ↪i_newInfectious, i_firstOrderDelay],
    O = [o_deaths, o_newInfectious, o_aging,
    ↪o_firstOrderDelay, o_deaths, o_newInfectious, o_aging,
    ↪o_firstOrderDelay, o_deaths, o_newInfectious,
    ↪o_firstOrderDelay],
    V = [v_births, v_newInfectious, v_newInfectious,
    ↪v_newInfectious, v_firstOrderDelay, v_firstOrderDelay,
    ↪v_firstOrderDelay, v_aging, v_aging, v_deaths, v_deaths,
    ↪v_deaths],
    LV = [lv_deaths, lv_newInfectious, lv_aging,
    ↪lv_firstOrderDelay, lv_deaths, lv_newInfectious, lv_aging,
    ↪lv_firstOrderDelay, lv_deaths, lv_newInfectious,
    ↪lv_firstOrderDelay],
    LSV = [lsv_N_births, lsv_NI_newInfectious,
    ↪lsv_NI_newInfectious, lsv_NI_newInfectious,
    ↪lsv_NS_newInfectious, lsv_NS_newInfectious,
    ↪lsv_NS_newInfectious, lsv_NI_newInfectious,
    ↪lsv_NI_newInfectious, lsv_NI_newInfectious,
    ↪lsv_NS_newInfectious, lsv_NS_newInfectious,
    ↪lsv_NS_newInfectious, lsv_NI_newInfectious,
    ↪lsv_NI_newInfectious, lsv_NI_newInfectious,
    ↪lsv_NS_newInfectious, lsv_NS_newInfectious,
    ↪lsv_NS_newInfectious],
    Name = name -> nothing
)

@assert is_natural(t_age_strata)
Graph_typed(t_age_strata)
```

[26]:

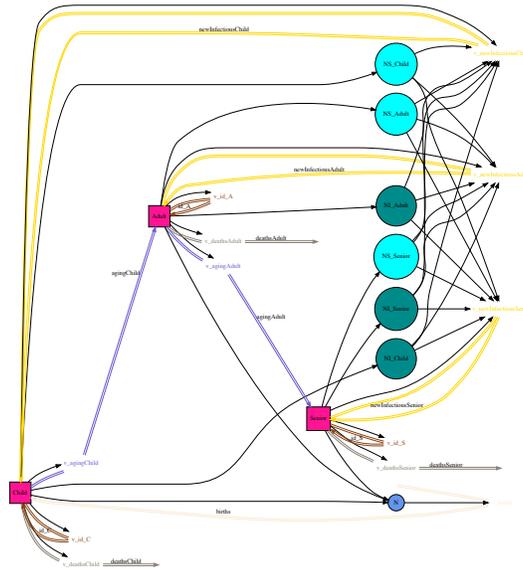

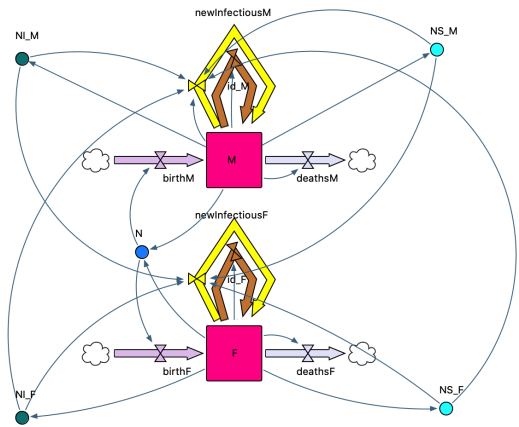

**Define the sex strata system structure diagram by two sub-groups: Female and Male**

[27]:
```
S_sex_strata=StockAndFlowStructure(
    (:F=>((:birthsF,:newInfectiousF,:id_F),(:deathsF,:
 ↪newInfectiousF,:id_F),(:v_deathsF,:v_newInfectiousF,:
 ↪v_idF),(:NI_F,:NS_F,:N)),
```

21

```
        :M=>((:birthsM,:newInfectiousM,:id_M),(:deathsM,:
    ↪newInfectiousM,:id_M),(:v_deathsM,:v_newInfectiousM,:
    ↪v_idM),(:NI_M,:NS_M,:N))),
    (:birthsF=>:v_birthsF,:birthsM=>:v_birthsM,:
    ↪newInfectiousF=>:v_newInfectiousF,:newInfectiousM=>:
    ↪v_newInfectiousM,:id_F=>:v_idF,:id_M=>:v_idM,
        :deathsF=>:v_deathsF,:deathsM=>:v_deathsM),
    (:N=>(:v_birthsF,:v_birthsM), :NI_F=>(:v_newInfectiousF,:
    ↪v_newInfectiousM), :NS_F=>(:v_newInfectiousF,:
    ↪v_newInfectiousM),
                    :NI_M=>(:v_newInfectiousF,:
    ↪v_newInfectiousM), :NS_M=>(:v_newInfectiousF,:
    ↪v_newInfectiousM))
)

t_sex_strata=ACSetTransformation(S_sex_strata, S_type,
    S = [s,s],
    SV = [sv_N, sv_NI, sv_NS, sv_NI, sv_NS],
    LS = [lsn_NI, lsn_NS, lsn_N, lsn_NI, lsn_NS, lsn_N],
    F = [f_births, f_births, f_newInfectious, f_newInfectious,
    ↪f_firstOrderDelay, f_firstOrderDelay, f_deaths, f_deaths],
    I = [i_births, i_newInfectious, i_firstOrderDelay,
    ↪i_births, i_newInfectious, i_firstOrderDelay],
    O = [o_deaths, o_newInfectious, o_firstOrderDelay,
    ↪o_deaths, o_newInfectious, o_firstOrderDelay],
    V = [v_births, v_births, v_newInfectious, v_newInfectious,
    ↪v_firstOrderDelay, v_firstOrderDelay, v_deaths, v_deaths],
    LV = [lv_deaths, lv_newInfectious, lv_firstOrderDelay,
    ↪lv_deaths, lv_newInfectious, lv_firstOrderDelay],
    LSV = [lsv_N_births, lsv_N_births, lsv_NI_newInfectious,
    ↪lsv_NI_newInfectious, lsv_NS_newInfectious,
    ↪lsv_NS_newInfectious, lsv_NI_newInfectious,
    ↪lsv_NI_newInfectious, lsv_NS_newInfectious,
    ↪lsv_NS_newInfectious],
    Name = name -> nothing
)

@assert is_natural(t_sex_strata)
Graph_typed(t_sex_strata)
```

[27]:

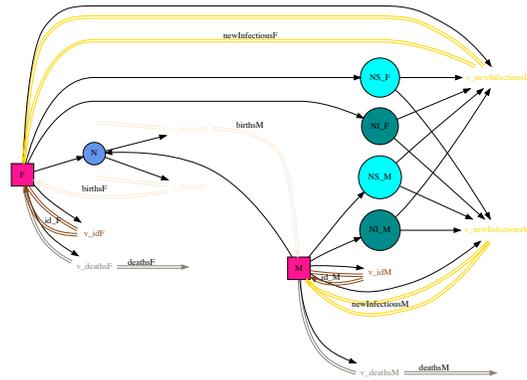

### 3.1.4 Build the stratified models

Firstly, we define the two functions of generating the stratified models and typed stratified models:

```
[28]: stratify(typed_models...) =
        ob(pullback(collect(typed_models)))

      typed_stratify(typed_models...) =
        compose(legs(pullback(collect(typed_models)))[1],
        typed_models[1]);
```

**Build the age stratified models**   The SEIR age stratified model:

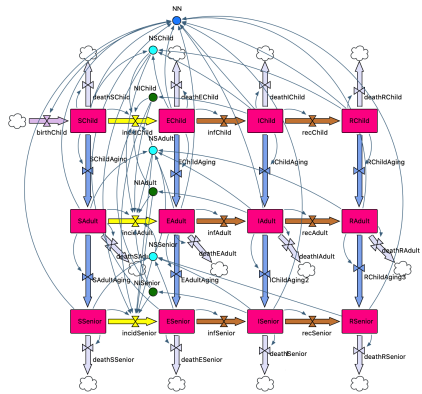

23

```
[29]: seir_age = stratify(t_seir, t_age_strata)
      Graph_typed(typed_stratify(t_seir, t_age_strata))
```

[29]:

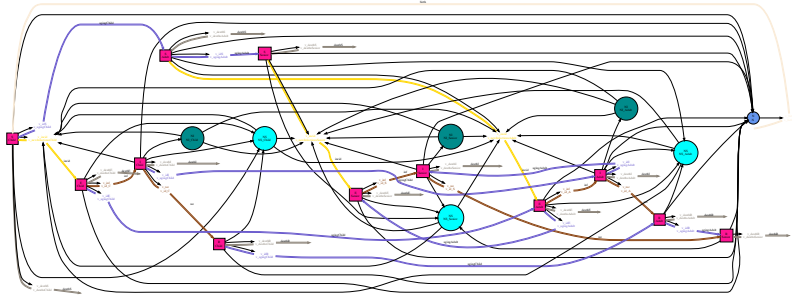

The SIS age stratified model:

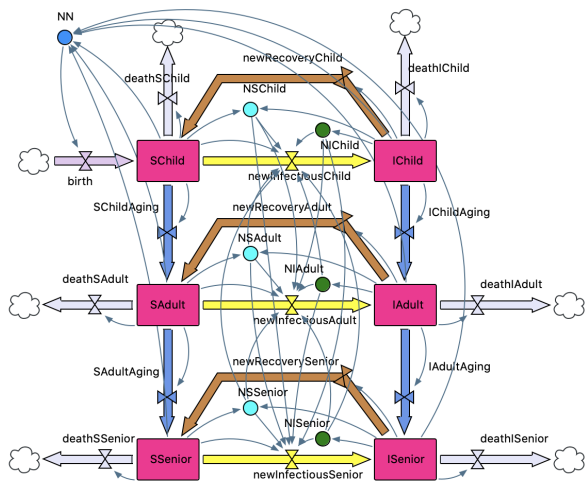

```
[30]: sis_age = stratify(t_sis, t_age_strata)
      Graph_typed(typed_stratify(t_sis, t_age_strata))
```

[30]:

24

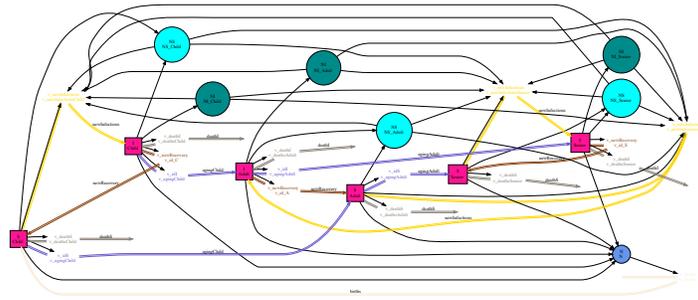

**Build the sex stratified models** The SEIR sex stratified model:

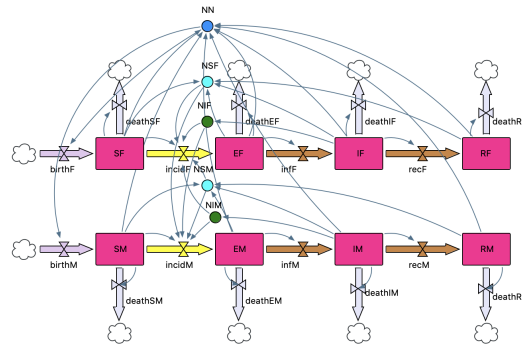

```
[31]: seir_sex = stratify(t_seir, t_sex_strata);
      Graph_typed(typed_stratify(t_seir, t_sex_strata))
```

[31]:

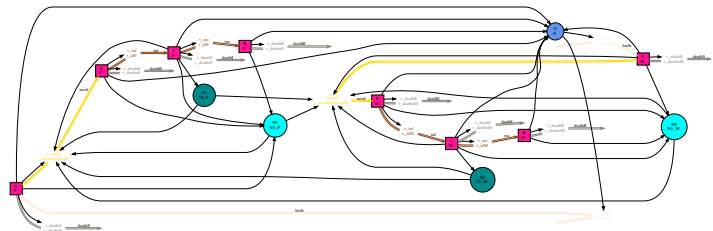

The SIS sex stratified model:

25

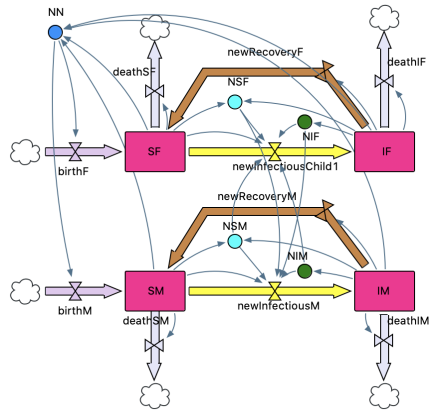

```
[32]: sis_sex = stratify(t_sis, t_sex_strata);
      Graph_typed(typed_stratify(t_sis, t_sex_strata))
```

[32]:

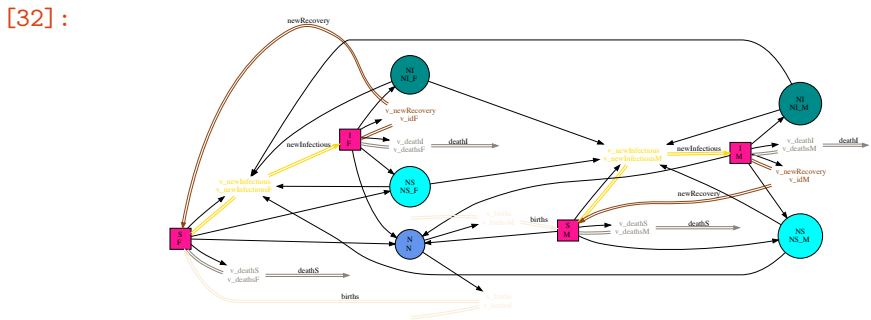

## 3.2 Build the stratified model with multiple strata (e.g., both age and sex strata)

We can build the stratified model with multiple dimensions (e.g., by age and sex) by calculating the pullback of the diseease model and multiple strata models. We create an example to stratify the SEIR model by both sex and age dimensions in this project. It is notable that when combine multiple strata model together to calculate the pullback, we may need to add some extra components to the strata models to accomodate to the other strata models. For example, we need to add an id flow of aging of each stock to the sex strata model to keep the aging flows in the stratified results.

26

```
[33]: S_sex_strata_withAge=StockAndFlowStructure(
      (:F=>((:birthsF,:newInfectiousF,:id_F,:agingF),(:
      ↪deathsF,:newInfectiousF,:id_F,:agingF),(:v_deathsF,:
      ↪v_newInfectiousF,:v_idF,:v_agingF),(:NI_F,:NS_F,:N)),
         :M=>((:birthsM,:newInfectiousM,:id_M,:agingM),(:
      ↪deathsM,:newInfectiousM,:id_M,:agingM),(:v_deathsM,:
      ↪v_newInfectiousM,:v_idM,:v_agingM),(:NI_M,:NS_M,:N))),
      (:birthsF=>:v_birthsF,:birthsM=>:v_birthsM,:
      ↪newInfectiousF=>:v_newInfectiousF,:newInfectiousM=>:
      ↪v_newInfectiousM,:id_F=>:v_idF,:id_M=>:v_idM,
         :deathsF=>:v_deathsF,:deathsM=>:v_deathsM,:agingF=>:
      ↪v_agingF,:agingM=>:v_agingM),
      (:N=>(:v_birthsF,:v_birthsM), :NI_F=>(:v_newInfectiousF,:
      ↪v_newInfectiousM), :NS_F=>(:v_newInfectiousF,:
      ↪v_newInfectiousM),
                  :NI_M=>(:v_newInfectiousF,:
      ↪v_newInfectiousM), :NS_M=>(:v_newInfectiousF,:
      ↪v_newInfectiousM))
)

t_sex_strata_withAge=ACSetTransformation(S_sex_strata_withAge,
      ↪S_type,
   S = [s,s],
   SV = [sv_N, sv_NI, sv_NS, sv_NI, sv_NS],
   LS = [lsn_NI, lsn_NS, lsn_N, lsn_NI, lsn_NS, lsn_N],
   F = [f_births, f_births, f_newInfectious, f_newInfectious,
      ↪f_firstOrderDelay, f_firstOrderDelay, f_deaths, f_deaths,
      ↪f_aging, f_aging],
   I = [i_births, i_newInfectious, i_firstOrderDelay,
      ↪i_aging, i_births, i_newInfectious, i_firstOrderDelay,
      ↪i_aging],
   O = [o_deaths, o_newInfectious, o_firstOrderDelay,
      ↪o_aging, o_deaths, o_newInfectious, o_firstOrderDelay,
      ↪o_aging],
   V = [v_births, v_births, v_newInfectious, v_newInfectious,
      ↪v_firstOrderDelay, v_firstOrderDelay, v_deaths, v_deaths,
      ↪v_aging, v_aging],
   LV = [lv_deaths, lv_newInfectious, lv_firstOrderDelay,
      ↪lv_aging, lv_deaths, lv_newInfectious, lv_firstOrderDelay,
      ↪lv_aging],
```

```
  LSV = [lsv_N_births, lsv_N_births, lsv_NI_newInfectious,
  ↪lsv_NI_newInfectious, lsv_NS_newInfectious,
  ↪lsv_NS_newInfectious, lsv_NI_newInfectious,
  ↪lsv_NI_newInfectious, lsv_NS_newInfectious,
  ↪lsv_NS_newInfectious],
  Name = name -> nothing
)

@assert is_natural(t_sex_strata_withAge)
Graph_typed(t_sex_strata_withAge)
```

[33]:

[34]: `seir_sex_age = stratify(t_seir, t_sex_strata_withAge,`
      `↪t_age_strata)`

```
Graph_typed(typed_stratify(t_seir, t_sex_strata_withAge,␣
 ↪t_age_strata))
```

[34]:

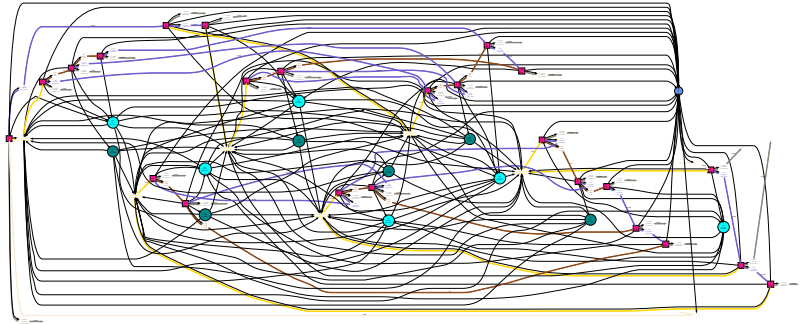

## 3.3 Solve the stratified model

We can assign the functions to the auxiliary variables and convert stratified system structure diagrams to stock-flow diagrams. Then, we can map the stock-flow diagrams to the ODEs, and solve it.

[35]:
```
sis_sex_flatten=rebuildSystemStructureByFlattenSymbols(sis_sex);
 ↪
```

Define the function for each auxiliary variable, and map to stock-flow diagram:

[36]:
```
f_birthF(u,uN,p,t)=p.μF*uN.NN(u,t)
f_birthM(u,uN,p,t)=p.μM*uN.NN(u,t)

f_infF(u,uN,p,t)= p.βF*u.SF*(p.ff*uN.NINI_F(u,t)/uN.
 ↪NSNS_F(u,t)+p.fm*uN.NINI_M(u,t)/uN.NSNS_M(u,t))
f_infM(u,uN,p,t)= p.βM*u.SM*(p.mf*uN.NINI_F(u,t)/uN.
 ↪NSNS_F(u,t)+p.mm*uN.NINI_M(u,t)/uN.NSNS_M(u,t))

f_recF(u,uN,p,t)=u.IF/p.trecoveryF
f_recM(u,uN,p,t)=u.IM/p.trecoveryM

f_deathSF(u,uN,p,t)=u.SF*p.δF
f_deathIF(u,uN,p,t)=u.IF*p.δF
f_deathSM(u,uN,p,t)=u.SM*p.δM
f_deathIM(u,uN,p,t)=u.IM*p.δM
```

```
v_sis_sex=(:v_birthsv_birthsF=>f_birthF, :
 ↪v_birthsv_birthsM=>f_birthM, :
 ↪v_newInfectiousv_newInfectiousF=>f_infF, :
 ↪v_newInfectiousv_newInfectiousM=>f_infM,
:v_newRecoveryv_idF=>f_recF, :v_newRecoveryv_idM=>f_recM, :
 ↪v_deathSv_deathsF=>f_deathSF, :
 ↪v_deathIv_deathsF=>f_deathIF, :
 ↪v_deathSv_deathsM=>f_deathSM, :
 ↪v_deathIv_deathsM=>f_deathIM);
```

Convert the system structure diagram to the stock-flow diagram:

```
[37]: sis_sex_SF=convertSystemStructureToStockFlow(sis_sex_flatten,v_sis_sex);
 ↪
```

Solve the ODEs of the converted stock-flow diagram:

```
[38]: p_sis_sex = LVector(
         βF=0.5, βM=0.6, μM=0, μF=0.03/365/2.0, δM=0.05/365/2.0,␣
      ↪δF=0.01/365/2.0, trecoveryM=5.0, trecoveryF=5.0,
         ff=0.5, fm=0.5, mf=0.5, mm=0.5
      )
      u0_sis_sex = LVector(
         SM=5400.0, SF=4600.0, IM=10.0, IF=1.0
      )

      prob_sis_sex =␣
       ↪ODEProblem(vectorfield(sis_sex_SF),u0_sis_sex,(0.0,50.
       ↪0),p_sis_sex);
      sol_sis_sex = solve(prob_sis_sex,Tsit5(),abstol=1e-8);
      plot(sol_sis_sex)
```

[38]:

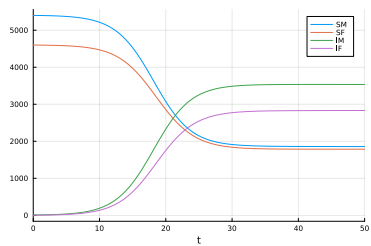


\end{document}